\documentclass[twocolumn]{aastex631}

\usepackage{ gensymb }
\usepackage{textcomp}
\usepackage{mathtools}
\usepackage{booktabs}
\usepackage{amsmath}
\usepackage{comment}
\usepackage{xcolor}

\shorttitle{UV--Optical Structural Evolution of LAEs at $z=2$--$7$}
\shortauthors{Laishram et al.}

\begin{document}

\title{Rest-Frame UV and Optical Structural Evolution of Narrowband-Selected Ly$\alpha$ Emitters at $z=2$--$7$ with JWST/NIRCam}

\correspondingauthor{Ronaldo Laishram}
\email{ronaldo.laishram@nao.ac.jp}

\author[0000-0002-0322-6131]{Ronaldo Laishram}
\affiliation{National Astronomical Observatory of Japan,
2-21-1 Osawa, Mitaka, Tokyo 181-8588, Japan}

\author[0000-0002-3801-434X]{Haruka Kusakabe}
\affiliation{Department of General Systems Studies, Graduate School of Arts and Sciences, The University of Tokyo, 3-8-1 Komaba, Meguro-ku, Tokyo, 153-8902, Japan}

\author[0000-0003-3214-9128]{Satoshi Kikuta}
\affiliation{Department of Regional Promotion, Nara Prefectural University, 10 Funahashicho, Nara, Nara, 630-8258, Japan}

\author[0009-0001-9612-1223]{Shunta Shimizu}
\affiliation{Department of Astronomy, School of Science, The University of Tokyo, 7-3-1, Hongo, Bunkyo, Tokyo 113-0033, Japan}

\author[0000-0002-0479-3699]{Yusei Koyama}
\affiliation{National Astronomical Observatory of Japan,
2-21-1 Osawa, Mitaka, Tokyo 181-8588, Japan}
\affiliation{Department of Astronomical Science,
The Graduate University for Advanced Studies,
2-21-1 Osawa, Mitaka,
Tokyo 181-8588, Japan}

\author[0000-0002-2993-1576]{Tadayuki Kodama}
\affiliation{Astronomical Institute, Tohoku University, 6-3, Aramaki, Aoba, Sendai, Miyagi 980-8578, Japan}

\author[0000-0002-3301-3321]{Michaela Hirschmann}
\affiliation{Institute of Physics, Laboratory for Galaxy Evolution, EPFL, Observatoire de Sauverny, Chemin Pegasi 51, CH-1290 Versoix, Switzerland}

\author[0009-0005-3133-1157]{Greta Toni}
\affiliation{Dipartimento di Fisica e Astronomia ``A. Righi'', Alma Mater Studiorum Universit\`a di Bologna, via Gobetti 93/2, 40129 Bologna, Italy}
\affiliation{INAF--Osservatorio di Astrofisica e Scienza dello Spazio di Bologna, via Gobetti 93/3, 40129 Bologna, Italy}
\affiliation{Zentrum f\"ur Astronomie, Universit\"at Heidelberg, Philosophenweg 12, 69120 Heidelberg, Germany}

\author[0009-0000-0272-5468]{Carter Flayhart}
\affiliation{Laboratory for Multiwavelength Astrophysics, School of Physics and Astronomy, Rochester Institute of Technology, 84 Lomb Memorial Drive, Rochester, NY 14623, USA}

\author[0000-0003-2716-8332]{Rasha M. Samir}
\affiliation{National Research Institute of Astronomy and Geophysics (NRIAG), Cairo 11421, Egypt}

\author[0009-0009-3404-5673]{Novan Saputra Haryana}
\affiliation{Astronomical Institute, Tohoku University, 6-3, Aramaki, Aoba, Sendai, Miyagi 980-8578, Japan}

\begin{abstract}
We present a systematic study of the rest-frame ultraviolet (UV) and optical morphologies of Ly$\alpha$ emitters (LAEs) at $z \simeq 2$--$7$ from the SILVERRUSH multi-narrowband catalog, using JWST/NIRCam imaging from COSMOS-Web. We fit S\'ersic profiles in four NIRCam filters (F115W, F150W, F277W, F444W) for LAEs spanning $\sim$2.2~Gyr of cosmic time. LAE sizes show a mild increase from rest-frame UV to optical wavelengths, with a median optical-to-UV size ratio of $R_{e,\mathrm{opt}}/R_{e,\mathrm{UV}} = 1.14^{+0.01}_{-0.02}$, indicating a small morphological $K$-correction; this offset is significant at $z \leq$ $5.7$ but not at $z =$ $6.6$. The rest-frame optical sizes follow a power-law evolution, given by $R_e \propto (1+z)^{-0.94 \pm 0.09}$, decreasing from $0.91^{+0.07}_{-0.07}$~kpc at $z =$ $2.2$ to $\sim$0.5~kpc at $z \gtrsim 5$, in close agreement with a recent JWST study of spectroscopically confirmed LAEs. The rest-frame optical size--mass relation slopes are broadly consistent with those of the general star-forming galaxy (SFG) population, while LAEs lie below it in normalization over $z \sim 3$--$6$. More compact LAEs tend to exhibit higher Ly$\alpha$ equivalent widths, a trend driven mainly by the most compact, marginally resolved sources. These compact systems also occupy elevated SFR surface-density regimes, consistent with a scenario in which concentrated star formation may facilitate Ly$\alpha$ escape. No significant correlation of LAE rest-frame optical size with projected large-scale environment is found over $z \simeq 2$--$7$, suggesting that LAE morphology may be governed primarily by internal galaxy properties, except possibly in the most extreme overdensities.
\end{abstract}


\section{Introduction} \label{sec:intro}

The structural properties of galaxies encode the physical processes governing their formation and assembly. At rest-frame ultraviolet (UV) wavelengths, galaxy morphology traces the spatial distribution of recent star formation, while rest-frame optical light is sensitive to the underlying stellar mass distribution built up over longer timescales \citep{Papovich_et_al_2005, Bond_et_al_2011}. Disentangling these two components is essential for understanding how galaxies grow, yet at $z \gtrsim 3$ this decomposition has remained observationally challenging. Prior to the James Webb Space Telescope (JWST), high-resolution rest-frame optical morphology at $z \gtrsim 3$ was difficult to measure for large samples, and most resolved morphological studies were restricted to the rest-frame UV. JWST/NIRCam now enables resolved measurements of both rest-frame UV and optical structures for large samples of high-redshift galaxies.

Lyman-$\alpha$ emitters (LAEs) provide large, narrowband-selected samples of line-emitting star-forming galaxies at $z \gtrsim 2$. Selected by their strong Ly$\alpha$ emission (rest-frame equivalent width EW$_0 \gtrsim 20$--$25$~\AA), such LAEs are typically characterized by low stellar masses ($M_\ast \sim 10^{8}$--$10^{9}~M_\odot$) and modest star formation rates (SFR $\sim 1$--$10~M_\odot$~yr$^{-1}$) \citep{Ouchi_et_al_2020, Kusakabe_et_al_2015, Kusakabe_et_al_2018}. Deeper spectroscopic surveys with MUSE, together with JWST follow-up, extend the LAE population to fainter, lower-mass systems \citep{Bacon_et_al_2023}, with some reaching stellar masses as low as $M_\ast \sim 10^{6}~M_\odot$ and more than half below $10^{8}~M_\odot$ \citep{Goovaerts_et_al_2024}. These properties make LAEs efficient probes of the low-mass star-forming galaxy population throughout the first few billion years of cosmic history \citep{Umeda_et_al_2025}. LAEs are generally dust-poor, though recent studies reveal considerable diversity in stellar age and star-formation history, from young ($<$100~Myr) bursty systems to older systems with more extended or complex star-formation histories \citep{Kusakabe_et_al_2015, Shimizu_et_al_2025, Firestone_et_al_2025}. Studies of faint, spectroscopically confirmed LAEs further reveal low gas-phase metallicities and hard ionizing radiation fields \citep{Nakajima_et_al_2018, Maseda_et_al_2023}. Moreover, the resonant scattering of Ly$\alpha$ photons by neutral hydrogen means that the detectability of Ly$\alpha$ emission is intimately connected to the geometry, kinematics, and dust content of the interstellar and circumgalactic medium, establishing a physical connection between galaxy structure and Ly$\alpha$ escape \citep{PaulinoAfonso_et_al_2018}.

Extensive efforts with the Hubble Space Telescope (HST) have established the rest-frame UV morphological properties of LAEs across $z \sim 2$--$6$. These studies consistently reveal compact systems with effective radii $r_e \sim 1$~kpc that exhibit little to no size evolution over this $\sim$3~Gyr period \citep{Taniguchi_et_al_2009, Gronwall_et_al_2011, Bond_et_al_2012, Malhotra_et_al_2012, Kobayashi_et_al_2016, PaulinoAfonso_et_al_2018}. \citet{PaulinoAfonso_et_al_2018} performed a statistical study with $\sim$4000 LAEs selected from 16 narrow- and medium-band filters in the COSMOS field, finding a size evolution of $r_e \propto (1+z)^{-0.21 \pm 0.22}$, consistent with no growth, though steeper rest-UV evolution has since been reported by other studies (see below). They further found that LAEs with higher Ly$\alpha$ equivalent widths tend to be smaller and more concentrated, suggesting that compact rest-UV morphology may be associated with conditions that favor Ly$\alpha$ escape. At $z \lesssim 2$, LAEs are a factor of $\sim$2--4 smaller than typical star-forming galaxies (SFGs), but the two populations converge in size at $z \gtrsim 5$ \citep{PaulinoAfonso_et_al_2018}. Gravitationally lensed LAEs at $z = 1.7$--$3.3$ reveal even more compact structures with a mean circularized effective radius $r_{e,\rm circ} = 170 \pm 140$~pc, underscoring the intrinsically small scale of UV-emitting regions in LAEs \citep{Kim_et_al_2026}. A larger sample of lensed Ly$\alpha$ emitters at $z = 2.9$--$6.7$, characterized using half-light and 90\%-light radii ($r_{50}$ and $r_{90}$), likewise reveals compact rest-UV sizes \citep{Claeyssens_et_al_2022}. However, a steeper size evolution ($r_{e,\rm circ} \propto (1+z)^{-1.37 \pm 0.65}$) has been reported when accounting for luminosity-dependent selection effects \citep{Shibuya_et_al_2019}, highlighting that the inferred evolution remains sensitive to sample selection and completeness.

JWST has enabled measurements of LAE structure at rest-frame optical wavelengths at $z > 3$. Recent studies indicate that LAEs remain compact in the rest-frame optical and that their size evolution and size--mass relation indicate structural growth toward lower redshift \citep{Song_et_al_2026, Shimizu_et_al_2025}. In particular, \citet{Song_et_al_2026} analyzed a large spectroscopically confirmed LAE sample at $3 \lesssim z < 7$, finding mild rest-frame optical size evolution and broadly similar rest-UV and rest-optical sizes. A JWST and HST study of narrowband-selected LAEs at $z = 2.2$--$6.6$ has likewise found compact rest-optical morphologies and suggested that LAEs are more compact than typical SFGs at lower redshifts but approach the SFG size--mass relation toward earlier cosmic times \citep{Shimizu_et_al_2025}. Using COSMOS-Web imaging at rest-frame $\sim$8000~\AA, \citet{Im_et_al_2026} found that narrowband-selected LAEs at $z = 2.4$--$4.5$ are smaller than typical SFGs of similar stellar mass. These results align with recent JWST studies of the general SFG population, which show that rest-frame UV and optical sizes are often comparable on average, while low-mass galaxies may exhibit stronger wavelength-dependent size differences at $z \gtrsim 5$ \citep{Ono_et_al_2024, Ward_et_al_2024, Allen_et_al_2025, Yang_et_al_2025, Yang_et_al_2026}.

In high-resolution zoom-in simulations at $z \gtrsim 5$, galaxies are more extended in rest-frame optical than in rest-frame UV light, with UV morphologies dominated by a few young stellar clumps \citep{Ma_et_al_2018}, and simulated galaxy sizes generally decrease toward higher redshift \citep{Ma_et_al_2018, LaChance_et_al_2025}. Simulations also predict that dust concentrated in galaxy cores can substantially increase observed rest-UV sizes relative to intrinsic sizes at $z \geq 5$ \citep{Roper_et_al_2022, Marshall_et_al_2022}. The smaller rest-frame optical sizes of LAEs relative to typical SFGs, and the decrease of this size difference toward higher redshift, are qualitatively reproduced in the Horizon Run 5 simulation \citep{Im_et_al_2026}.

However, several questions remain open. The comparison between rest-frame UV and optical structures has not been systematically explored across the full redshift range $z \sim 2$--$7$ using a single, uniformly selected sample with consistent methodology. The role of environment in shaping LAE structure at rest-optical wavelengths requires investigation across multiple redshifts and the full range of local densities. It also remains unclear whether the negligible UV--optical color gradients reported by \citet{Song_et_al_2026} persist across all redshifts, including $z \sim 2$ and within narrow redshift windows. Existing JWST measurements of LAE rest-frame optical morphology are based on spectroscopically confirmed samples \citep{Song_et_al_2026}, on a narrowband-selected sample analyzed primarily for the Ly$\alpha$ escape fraction \citep{Shimizu_et_al_2025}, or on narrowband-selected LAEs at a single rest-frame wavelength over $z = 2.4$--$4.5$ \citep{Im_et_al_2026}.

In this paper, we study the rest-frame UV and optical morphologies of LAEs spanning $z \sim 2$--$7$, selected from the SILVERRUSH narrowband survey \citep{Ouchi_et_al_2018, Kikuta_et_al_2023} and observed with JWST/NIRCam imaging from COSMOS-Web \citep{Casey_et_al_2023}. We measure S\'ersic profile parameters \citep{Sersic_1968} in rest-frame UV and optical bands, enabling a direct comparison between the rest-frame UV and optical light distributions. We analyze 770 narrowband-selected LAEs in six narrowband redshift windows and derive size evolution, the rest-frame optical size--mass relation, wavelength-dependent morphology, and environmental trends uniformly across this redshift range. Throughout this work, we adopt a flat $\Lambda$CDM cosmology with $\Omega_m = 0.3$, $\Omega_\Lambda = 0.7$, and $H_0 = 70$~km~s$^{-1}$~Mpc$^{-1}$, and quote magnitudes in the AB system \citep{Oke_Gunn_1983}.

\section{Data} \label{sec:data}

\subsection{The Parent LAE Sample} \label{sec:lae_sample}

Our parent sample of Ly$\alpha$ emitters is drawn from the large-scale LAE catalog presented by \citet{Kikuta_et_al_2023}, constructed using deep narrowband (NB) imaging data from the Hyper Suprime-Cam Subaru Strategic Program (HSC-SSP; \citealt{Aihara_et_al_2022}) and the Cosmic HydrOgen Reionization Unveiled with Subaru (CHORUS) survey \citep{Inoue_et_al_2020}. The HSC-SSP is a multi-tier wide-field imaging survey utilizing the 1.77~deg$^2$ field-of-view of HSC on the 8.2~m Subaru Telescope \citep{Aihara_et_al_2019, Aihara_et_al_2022}, with narrowband imaging available in its Deep and UltraDeep layers. Together, these data provide broadband ($grizy$) and narrowband photometry over an area of up to $\sim$25~deg$^2$; in this work, we use only LAEs within the 0.54~deg$^2$ COSMOS-Web NIRCam footprint (Section~\ref{sec:sample_construction}).

The catalog employs a series of narrowband filters to identify Ly$\alpha$ emission at discrete redshift windows spanning a broad range of cosmic epochs. The narrowband filters and their corresponding Ly$\alpha$ redshifts are: NB387 ($z = 2.18 \pm 0.023$), NB527 ($z = 3.33 \pm 0.032$), NB718 ($z = 4.90 \pm 0.046$), NB816 ($z = 5.72 \pm 0.046$), NB921 ($z = 6.58 \pm 0.056$), and NB973 ($z = 6.99 \pm 0.046$). LAE candidates were selected via narrowband excess (NB $-$ BB color) with $>5\sigma$ detection significance in each narrowband filter, combined with Lyman break color criteria using multi-band photometry to reject low-redshift interlopers \citep{Ouchi_et_al_2018, Kikuta_et_al_2023}. Source detection and photometry were performed on point spread function (PSF)-matched coadded images, with limiting magnitudes measured on a patch-by-patch basis to ensure uniform selection across the survey area. The catalog has undergone careful artifact masking and visual inspection to ensure high sample purity. Full details of the data reduction, photometric selection criteria, and masking procedures are provided in \citet{Kikuta_et_al_2023}.

\subsection{JWST COSMOS-Web Imaging Data} \label{sec:jwst_data}

We utilize deep near-infrared imaging from the COSMOS-Web program \citep{Casey_et_al_2023}, a JWST Cycle~1 treasury survey that mapped 0.54~deg$^2$ of the COSMOS field with NIRCam and 0.19~deg$^2$ with MIRI. The NIRCam observations employ four broadband filters (F115W, F150W, F277W, and F444W) \citep{Casey_et_al_2023, Shuntov_et_al_2025}. Observations were executed across multiple epochs between 2023 January and 2024 May. Full details of the NIRCam image processing and mosaic construction are provided by \citet{Franco_et_al_2024, Franco_et_al_2025}. We adopt the 30~mas pixel scale mosaics for our structural measurements to optimize spatial sampling of compact high-redshift sources.

The four NIRCam filters probe different rest-frame wavelengths depending on the redshift of the source. The rest-frame UV and optical filter assignments for each narrowband are listed in Table~\ref{tab:sample}. This redshift-dependent filter assignment provides an approximate separation between shorter-wavelength emission dominated by recent star formation and longer-wavelength emission that more closely traces the accumulated stellar mass distribution, although the exact rest-frame wavelength probed varies with redshift and the broadband measurements may include contributions from nebular emission lines at some redshifts.

\subsection{Photometric Catalog and Derived Physical Properties} \label{sec:phot_catalog}

Stellar masses and star formation rates are obtained from the COSMOS-Web photometric catalog presented by \citet{Shuntov_et_al_2025}, which provides multi-wavelength photometry and spectral energy distribution (SED) fitting results for all detected sources in the survey footprint. The catalog employs the LePhare SED-fitting code \citep{Arnouts_et_al_2002, Ilbert_et_al_2006} to derive photometric redshifts and physical parameters from composite photometry spanning 37~bands from $\sim$0.3~$\mu$m to $\sim$8~$\mu$m \citep{Shuntov_et_al_2025}.

Template fitting is based on \citet{Bruzual_Charlot_2003} stellar population synthesis models incorporating varied star formation histories, stellar ages, and dust attenuation prescriptions \citep{Calzetti_et_al_2000}, and assumes a \citet{Chabrier_2003} initial mass function (IMF). For each source, LePhare computes a probability density function of the photometric redshift; physical properties (stellar mass, SFR) are then calculated at the median photometric redshift. We do not refit the SEDs at the narrowband Ly$\alpha$ redshifts, but adopt the published LePhare stellar masses and SFRs evaluated at the median photometric redshift after applying our photometric-redshift consistency cut.

Stellar masses from different SED codes and template libraries can differ systematically, which affects the comparison of our size--mass results with those of \citet{Song_et_al_2026}.

\subsection{Spectroscopic Redshift Compilation} \label{sec:spec_catalog}

To validate photometric redshifts and assess the purity of our narrowband-selected LAE sample, we cross-reference our sources with the COSMOS spectroscopic redshift compilation of \citet{Khostovan_et_al_2026}. This compilation aggregates spectroscopic redshift measurements for galaxies from spectroscopic programs observed over two decades within a 10~deg$^2$ area centered on the COSMOS field, spanning redshifts from the local Universe to $z \sim 8$. We use quality flags $\geq 3$ (reliable or very secure redshifts) to identify spectroscopically confirmed galaxies within the transmission windows of each narrowband filter.

\subsection{Sample Construction} \label{sec:sample_construction}

To construct the morphological sample, we cross-match the parent LAE catalog with the COSMOS-Web NIRCam imaging footprint using a matching radius of 0\farcs5, yielding 1811 LAEs, then apply a two-stage cleaning procedure to ensure sample purity. First, we cross-reference all candidates with the COSMOS spectroscopic redshift compilation \citep{Khostovan_et_al_2026} using a matching radius of 0\farcs5 to identify and remove sources with reliable spectroscopic redshifts differing from the narrowband redshift by more than 0.5. In the final sample, 55 LAEs (7\%) have reliable spectroscopic redshifts, 54 of which lie within 0.1 of the narrowband redshift. Increasing the matching radius to 1\farcs0 does not change the final sample. Second, we apply photometric redshift quality cuts using the COSMOS-Web LePhare-derived redshifts \citep{Shuntov_et_al_2025}, retaining only sources whose narrowband Ly$\alpha$ redshift lies within the 68\% confidence interval of the photometric redshift, extended by $\pm 0.2$. This criterion is intended to reduce contamination from galaxies outside the targeted redshift ranges, although it may also exclude genuine LAEs with uncertain or template-dependent photometric redshifts. Most of the excluded sources ($85\%$) have photometric redshifts at $z < 2$ (median $z \approx 0.7$), inconsistent with the Ly$\alpha$ redshift window of their narrowband filter. These redshifts often coincide with those at which other strong emission lines fall within the corresponding narrowband filter ($74\%$ lie within $\Delta z = 0.2$ of such a redshift, compared with $9\%$ of the retained LAEs), suggesting that the photo-$z$ cut primarily removes low-redshift emission-line interlopers rather than sources scattered into the LAE selection by photometric noise. Because the narrowband Ly$\alpha$ redshift is more precise than the broadband photo-$z$, the photo-$z$ cut is used only to remove clear outliers, not to determine redshifts. For all LAEs in a given narrowband, we adopt the central Ly$\alpha$ redshift of that filter (Section~\ref{sec:lae_sample}) to convert angular sizes to physical sizes and to derive $M_{1500}$ and EW$_0$.

The final sample consists of 770 LAEs spanning $z=2.18$--$6.99$ after imposing $\log(M_\ast/M_\odot) \geq 8.0$, so that all narrowband redshift bins cover a similar stellar-mass range. Because LAEs are selected by Ly$\alpha$ narrowband excess, the sample completeness is governed by Ly$\alpha$ line flux and equivalent width thresholds of the narrowband survey \citep{Kikuta_et_al_2023}. The relevant selection limits in terms of EW$_0$ and $M_{\rm UV}$ have been characterized by \citet{Shimizu_et_al_2025}, who analyzed LAEs from the same \citet{Kikuta_et_al_2023} parent catalog. Table~\ref{tab:sample} summarizes the sample properties for each narrowband selection, including the number of sources, redshift, and JWST filter used for morphology. NB973 ($z = 6.99$, $N = 4$) is excluded from all quantitative analyses throughout this paper owing to its small sample size; the working analytical sample therefore comprises 766 LAEs.

\begin{deluxetable*}{lcccc}
\tablecaption{LAE Sample Properties by Redshift\label{tab:sample}}
\tablehead{
\colhead{Narrowband} & \colhead{Redshift} & \colhead{$N_{\rm LAE}$} & \colhead{JWST Filter (UV)} & \colhead{JWST Filter (Opt)}
}
\startdata
NB387 & 2.18 & 154 & F115W & F150W \\
NB527 & 3.33 & 360 & F115W & F277W \\
NB718 & 4.90 & 132 & F150W & F277W \\
NB816 & 5.72 & 69  & F150W & F277W \\
NB921 & 6.58 & 51  & F277W & F444W \\
NB973\tablenotemark{a} & 6.99 & 4   & F277W & F444W \\
\hline
Total & 2.18--6.99 & 770 & --- & --- \\
\enddata
\tablenotetext{a}{Small sample size ($N=4$); excluded from all quantitative analyses.}
\tablecomments{Sample properties for our LAE sample after quality cuts. Median sizes and size--mass fit parameters are listed in Table~\ref{tab:size_mass}.}
\end{deluxetable*}

\section{Methodology} \label{sec:analysis}

\subsection{Morphological Measurements} \label{sec:morph_measurements}

To determine the structural properties of the LAEs, we perform two-dimensional surface brightness profile fitting using {\tt GALFIT} \citep{Peng_et_al_2002, Peng_et_al_2010} in all four NIRCam filters (F115W, F150W, F277W, F444W), enabling measurements at both rest-frame UV and optical wavelengths. For each LAE, we create a cutout image from the JWST mosaic centered on the source coordinates of the parent LAE catalog \citep{Kikuta_et_al_2023}, which correspond to SExtractor centroids measured on the narrowband detection images. For the primary rest-frame optical analysis, we adopt the longer-wavelength filter at each redshift, corresponding to rest-frame wavelengths of $\lambda_{\rm rest} \simeq 4100$--$6400$\,\AA: F150W ($z=2.18$), F277W ($z=3.33$, $4.90$, $5.72$), and F444W ($z=6.58$, $6.99$). Rest-frame UV sizes are measured from the corresponding shorter-wavelength filters listed in Table~\ref{tab:sample}.

We model the galaxy light distribution using a single S\'ersic profile \citep{Sersic_1968}:
\begin{equation} \label{eq:sersic}
\Sigma(r) = \Sigma_e \exp \left[ -b_n \left( \left( \frac{r}{R_e} \right)^{1/n} - 1 \right) \right],
\end{equation}
where $R_e$ is the effective radius along the semimajor axis, $n$ is the S\'ersic index, and $\Sigma_e$ is the surface brightness at $R_e$. The constant $b_n$ is coupled to $n$ such that $\Gamma(2n) = 2\gamma(2n, b_n)$, ensuring that $R_e$ encloses half of the total light.

Accurate PSF modeling is critical for reliable structural measurements of high-redshift galaxies, which are often comparable in angular size to the PSF \citep[e.g.,][]{Kim_et_al_2026, Song_et_al_2026}. We construct one empirical PSF model per NIRCam filter by stacking isolated, unsaturated stars identified within the COSMOS-Web field. The resulting full widths at half maximum (FWHMs) of our PSF models are 0\farcs057, 0\farcs060, 0\farcs121, and 0\farcs158 for the F115W, F150W, F277W, and F444W filters, respectively.

During the fitting process, we allow the effective radius ($R_e$), S\'ersic index ($n$), axis ratio ($b/a$), position angle (PA), and total magnitude to vary as free parameters. We constrain the S\'ersic index to the range $0.3 < n < 8$ and the effective radius to $R_e < 50$~kpc to prevent unphysical solutions, following standard practices in high-redshift galaxy morphology \citep[e.g.,][]{van_der_Wel_et_al_2012, Shibuya_et_al_2019}.

\subsection{UV Continuum and Ly$\alpha$ Equivalent Width} \label{sec:uv_ew}

We derive the absolute UV magnitude ($M_{\rm UV}$) and rest-frame Ly$\alpha$ equivalent width (EW$_0$) for our sample following the methodology described in \citet{Shibuya_et_al_2018}.

\subsubsection{Absolute UV Magnitude} \label{sec:muv}

To estimate the rest-frame UV luminosity, we use the broadband filter closest to rest-frame $\sim$1500\,\AA\ that is uncontaminated by the Ly$\alpha$ line. We calculate the rest-frame wavelength probed by each filter as $\lambda_{\rm rest} = \lambda_{\rm obs} / (1+z)$. Following \citet{Shibuya_et_al_2018} and \citet{Umeda_et_al_2025}, we assume a flat UV continuum in frequency ($f_\nu = \mathrm{const}$, equivalent to $\beta = -2$ in the common $f_\lambda \propto \lambda^\beta$ convention), so no color correction is applied between the observed broadband and rest-frame 1500\,\AA, and we take $m_{1500} = m_{\rm obs}$. The absolute magnitude $M_{1500}$ is then derived using the luminosity distance $D_L(z)$ and the $K$-correction term $2.5\log_{10}(1+z)$, assuming our fiducial $\Lambda$CDM cosmology. Sources undetected in the broadband used for $M_{1500}$ are excluded from size--$M_{\rm UV}$ analyses; 753 of the 766 LAEs have reliable $M_{1500}$ measurements.

\subsubsection{Ly$\alpha$ Equivalent Width} \label{sec:ew}

The rest-frame Ly$\alpha$ equivalent width (EW$_{0,{\rm Ly}\alpha}$) is calculated using the narrowband and broadband (BB) photometry, with the NB--BB pairings following \citet{Shibuya_et_al_2018} and \citet{Kikuta_et_al_2023}. We adopt the formalism of \citet{Shibuya_et_al_2018}, which accounts for the contributions of both the Ly$\alpha$ line and the UV continuum to the observed flux in each filter bandpass. Assuming a spectrally unresolved Ly$\alpha$ line and a flat UV continuum in frequency ($f_\nu = \mathrm{const}$, i.e., $\beta = -2$; \citealt{Shibuya_et_al_2018, Umeda_et_al_2025}), the line flux ($f_l$) and continuum flux density ($f_c$) are derived by solving the system of linear equations relating the observed NB and BB flux densities.

For sources with broadband magnitudes fainter than the $1\sigma$ limiting magnitude, we adopt the $1\sigma$ limit for the calculation, yielding a lower bound on EW$_0$; this applies to 26 of the 766 LAEs, all at $z \geq 3.33$. All 766 LAEs are included in the size--EW correlation analysis, and excluding these 26 LAEs does not change the results of Section~\ref{sec:size_ew}. To estimate uncertainties on EW$_{0,{\rm Ly}\alpha}$, we perform a Monte Carlo simulation with $10{,}000$ realizations, perturbing the input fluxes according to their photometric errors and varying the UV continuum slope around $\beta = -2$ with a dispersion of $\sigma_\beta = 0.2$, following \citet{Shibuya_et_al_2018}.

\subsection{Local Environment Estimation} \label{sec:env_method}

We characterize the local LAE surface density using the fifth-nearest-neighbor overdensity, $\delta_\mathrm{NN}$ \citep{Ramella_et_al_2001, Cooper_et_al_2005, Darvish_et_al_2015}. For each LAE, $\delta_\mathrm{NN} = \Sigma_5/\langle\Sigma_5\rangle - 1$, where $\Sigma_5 = 5/(\pi d_5^2)$ is the projected surface density defined by the angular distance $d_5$ to the fifth nearest neighbor (converted to physical Mpc at the LAE's narrowband redshift), and $\langle\Sigma_5\rangle$ is the median surface density computed from the parent LAE catalog in the same narrowband redshift slice. The overdensity is evaluated from the full parent LAE catalog in each narrowband field independently, and therefore traces the projected LAE density rather than the full galaxy density field. As a secondary estimator, we also compute a Voronoi tessellation overdensity $\delta_\mathrm{Vor}$ for each LAE, in which the local surface density is estimated from the inverse area of the Voronoi cell constructed around each source in the same narrowband slice. Unlike $\delta_\mathrm{NN}$, the Voronoi estimator is non-parametric and does not require a choice of neighbor number \citep{Darvish_et_al_2015}, and comparing the two provides a check on whether our results depend on the adopted density estimator. We adopt $\delta_\mathrm{NN}$ as the primary estimator for consistency with \citet{Laishram_et_al_2026b}, allowing comparison with that study. Edge effects in the overdensity measurements are mitigated because the COSMOS-Web footprint is embedded within the larger SILVERRUSH/HSC survey area, ensuring that all LAEs have complete neighbor coverage for environment calculations \citep{Laishram_et_al_2026b}.

\section{Results} \label{sec:results}

For morphological analyses, we compare our measurements with those of \citet{Song_et_al_2026}, who analyzed 876 spectroscopically confirmed LAEs at $3 \lesssim z < 7$ using JWST/NIRCam imaging. Our uniformly narrowband-selected sample of 766 LAEs from COSMOS-Web provides an independent dataset with narrow redshift windows set by the NB filter widths, extends the baseline down to $z = 2.18$, and includes measurements of the local environment.

\subsection{Wavelength Dependence of Morphology} \label{sec:wavelength_dependence}

\begin{figure}
\includegraphics[width=0.99\columnwidth]{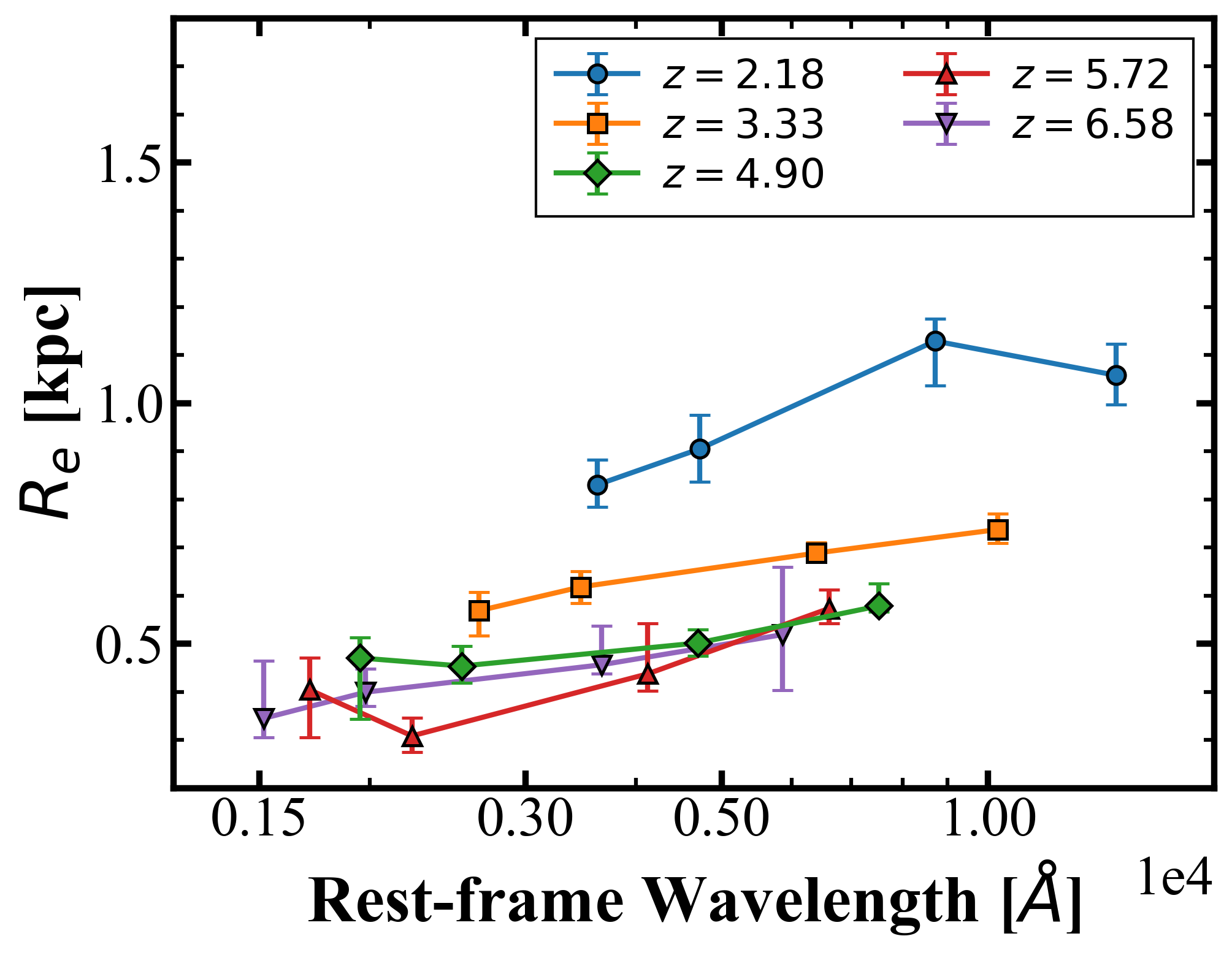}
\caption{Median effective radius $R_e$ as a function of rest-frame wavelength for LAEs at each narrowband redshift, measured in all four NIRCam filters. Error bars show 68\% bootstrap confidence intervals ($10{,}000$ iterations). NB973 ($z = 6.99$) is excluded ($N = 4$).}
\label{fig:size_vs_wavelength}
\end{figure}

We first examine how LAE sizes depend on the rest-frame wavelength of observation by measuring $R_e$ in all four NIRCam filters (F115W, F150W, F277W, F444W) at each redshift. Across the full redshift range, these filters probe rest-frame wavelengths from $\sim$1500~\AA\ to $\sim$14{,}000~\AA, although the exact coverage differs for each narrowband redshift. Figure~\ref{fig:size_vs_wavelength} shows the median $R_e$ as a function of rest-frame wavelength at each redshift.

LAE sizes generally show a mild increase toward longer rest-frame wavelengths, although the significance of this trend varies with redshift (Figure~\ref{fig:size_vs_wavelength}). For example, at $z = 3.33$ ($N = 360$), the median $R_e$ increases from $0.62^{+0.03}_{-0.03}$~kpc at rest-frame $\sim$3500~\AA\ to $0.74^{+0.03}_{-0.03}$~kpc at $\sim$10{,}300~\AA. Comparing shorter- and longer-wavelength measurements separated at $\lambda_\mathrm{rest} = 4000$~\AA\ (the `UV' band spans $\sim$2200--3600~\AA\ depending on redshift; Table~\ref{tab:sample}) for 766 LAEs, we find a median ratio of $R_{e,\mathrm{opt}}/R_{e,\mathrm{UV}} = 1.14^{+0.01}_{-0.02}$ ($\log(R_{e,\mathrm{opt}}/R_{e,\mathrm{UV}}) = 0.056^{+0.005}_{-0.006}$; Wilcoxon signed-rank $p = 6.3 \times 10^{-19}$). The offset is statistically significant at $z = 2.18$ ($p = 0.001$), $z = 3.33$ ($p = 4.4 \times 10^{-12}$), $z = 4.90$ ($p = 0.005$), and $z = 5.72$ ($p = 3.7 \times 10^{-5}$), but is not significant at $z = 6.58$ ($p = 0.46$); the per-narrowband ratios are listed in Table~\ref{tab:size_mass}. The offset reported here ($\sim$0.06~dex) is somewhat larger than the $\sim$0.03~dex UV--optical size difference found by \citet{Song_et_al_2026}, a difference that may in part reflect the broader wavelength baseline in the present study: \citet{Song_et_al_2026} compared sizes at rest-frame $\sim$2800~\AA\ and $\sim$5000~\AA, whereas our measurements span a wider, redshift-dependent range (Table~\ref{tab:sample}). We interpret this offset as a broadband morphological $K$-correction rather than solely a physical stellar-continuum color gradient. Residual PSF-model uncertainties, which have a larger relative impact in the longer-wavelength bands where sources are less well resolved, may also contribute to the observed offset. For the remainder of this paper, we adopt the rest-frame optical $R_e$ as the primary size measurement.

\subsection{Size Evolution} \label{sec:size_evolution}

\begin{figure*}
\includegraphics[width=\textwidth]{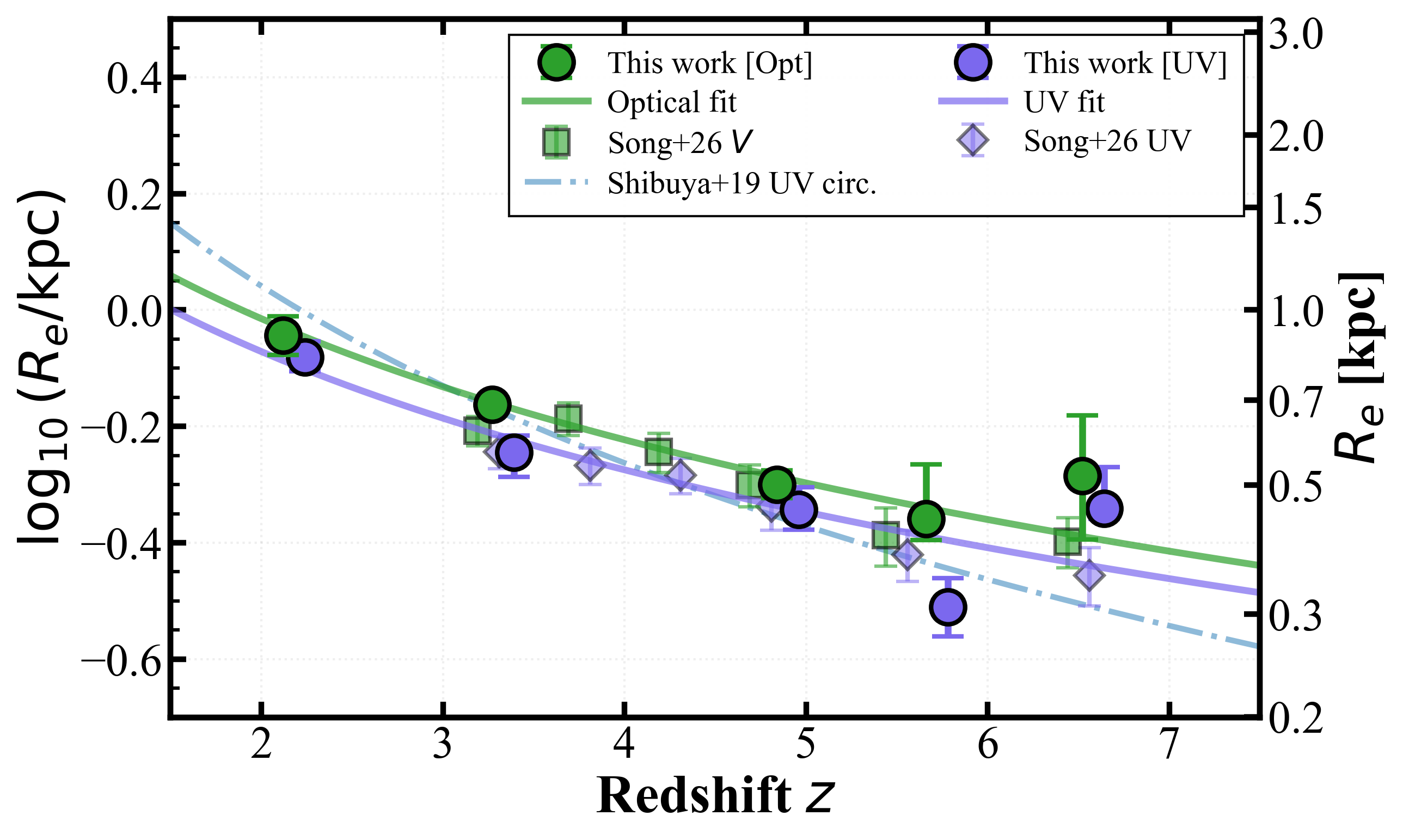}
\caption{Redshift evolution of the median effective radius for LAEs in the rest-frame optical (green circles) and rest-frame UV (purple circles). Error bars show 68\% bootstrap confidence intervals ($10{,}000$ iterations). Solid lines show the best-fit power laws $R_e = \alpha\,(1+z)^\beta$ for each band. Literature data are shown for comparison: \citet{Song_et_al_2026} $V$-band (green squares) and UV (purple diamonds); and the \citet{Shibuya_et_al_2019} UV power-law fit (blue dash-dotted line). Rest-frame optical and $V$-band data points are offset by $\Delta z = -0.06$, and rest-frame UV data points by $\Delta z = +0.06$, for visual clarity. The secondary $y$-axis shows $R_e$ in kpc.}
\label{fig:size_evolution}
\end{figure*}

We investigate the redshift evolution of LAE sizes by computing the median $R_e$ in each narrowband redshift window for both rest-frame optical and rest-frame UV. At each redshift, we estimate the median $\log_{10}(R_e/\mathrm{kpc})$ and its 68\% confidence interval from $10{,}000$ bootstrap iterations. The size-evolution results are presented in Figure~\ref{fig:size_evolution}; the per-narrowband median sizes are listed in Table~\ref{tab:size_mass}.

In the rest-frame optical, the median $R_e$ decreases from $0.91^{+0.07}_{-0.07}$~kpc at $z = 2.18$ to $0.44^{+0.10}_{-0.04}$~kpc at $z = 5.72$, and the median at $z = 6.58$ is consistent within uncertainties with the values at $z \simeq 4.9$--$5.7$. The rest-frame UV sizes follow a similar trend (Table~\ref{tab:size_mass}).

Following previous studies \citep{PaulinoAfonso_et_al_2018, Shibuya_et_al_2019, Song_et_al_2026}, we parameterize the size evolution as a power law:
\begin{equation} \label{eq:size_evolution}
R_e = \alpha\,(1+z)^\beta \quad [\mathrm{kpc}],
\end{equation}
where $\alpha$ and $\beta$ are free parameters fitted to the median $\log_{10}(R_e/\mathrm{kpc})$ values at each narrowband redshift, weighted by the bootstrap uncertainties. For the rest-frame optical, we obtain $(\alpha,\,\beta) = (2.71 \pm 0.35,\;-0.94 \pm 0.09)$, and for the rest-frame UV $(\alpha,\,\beta) = (2.32 \pm 0.78,\;-0.92 \pm 0.21)$. Both bands yield consistent power-law indices within uncertainties, suggesting no strong wavelength dependence in the inferred rate of size evolution.

Our rest-frame optical power-law index, $\beta_\mathrm{opt} = -0.94 \pm 0.09$, is consistent with the $V$-band result of \citet{Song_et_al_2026}, who found $\beta_V = -0.93 \pm 0.18$ for spectroscopically confirmed LAEs at $3 \lesssim z < 7$. The inclusion of the $z = 2.18$ bin extends the redshift baseline below the range covered by \citet{Song_et_al_2026}, providing additional leverage on the fitted slope; excluding this bin yields a similar slope, $\beta_\mathrm{opt} = -0.96 \pm 0.14$. The UV slope, $\beta_\mathrm{UV} = -0.92 \pm 0.21$, is likewise consistent with their UV result, $\beta_\mathrm{UV} = -0.91 \pm 0.10$; our adopted rest-frame UV wavelengths vary from $\sim$2200~\AA\ to $\sim$3600~\AA\ depending on redshift, compared to their fixed $\sim$2800~\AA\ definition.

The mild size evolution we find ($\beta \approx -0.9$) is intermediate between the nearly flat trend reported by \citet{PaulinoAfonso_et_al_2018} ($\beta = -0.21 \pm 0.22$; rest-frame UV, circularized radii, $z \sim 2$--$6$) and the steeper evolution found by \citet{Shibuya_et_al_2019} ($\beta = -1.37 \pm 0.65$; rest-frame UV, circularized radii, luminosity-matched). Several factors may contribute to the differences among these studies. First, the radius conventions differ: \citet{PaulinoAfonso_et_al_2018} and \citet{Shibuya_et_al_2019} report circularized radii ($r_{e,\mathrm{circ}} = R_e \sqrt{b/a}$), which are smaller than semimajor-axis radii by a factor of $\sqrt{b/a} \sim 0.7$ for typical LAE axis ratios ($b/a \sim 0.45$--$0.50$; \citealt{Song_et_al_2026}). This convention primarily affects the normalization rather than the fitted slope, as the LAE axis-ratio distribution shows no strong redshift evolution \citep{Song_et_al_2026}. Second, the \citet{PaulinoAfonso_et_al_2018} sample is restricted to sources brighter than $i_\mathrm{AB} < 25$ observed with HST/ACS F814W; this flux limit preferentially excludes compact, faint LAEs at higher redshift, which, given the size--luminosity relation, biases the high-redshift size distribution toward larger values and may flatten the inferred slope relative to deeper samples. Third, \citet{Shibuya_et_al_2019} applied a UV luminosity cut ($L_\mathrm{UV} = 0.12$--$1\,L^*_{z=3}$) to control for luminosity-dependent selection effects, and their steeper slope likely also reflects the different redshift baseline and sample selection relative to our study. Our results are in close agreement with the recent JWST-based measurements of \citet{Song_et_al_2026}, supporting a mild decrease in LAE sizes with increasing redshift over $z \sim 2$--$7$. Star-forming galaxies in COSMOS-Web show a somewhat steeper rest-frame optical size evolution at fixed stellar mass ($R_e \propto (1+z)^{-1.21 \pm 0.05}$; \citealt{Yang_et_al_2025}), and the size difference between LAEs and typical SFGs decreases toward higher redshift \citep{Im_et_al_2026}. UV-bright Lyman-break galaxies with $\log(M_\ast/M_\odot) > 9$ at $z = 3$--$5$ show a shallower rest-frame optical size evolution, $R_e \propto (1+z)^{-0.60 \pm 0.22}$, although with larger uncertainty \citep{Varadaraj_et_al_2024}.

The rest-frame optical S\'{e}rsic indices are generally low, with median $n$ ranging from $0.30$ to $1.17$. The lowest-redshift bin ($z \sim 2.2$) has a median $n \approx 1.17$, while the $z > 3$ bins have median values below unity, indicating sub-exponential or low-concentration light profiles rather than centrally concentrated morphologies. Higher median S\'ersic indices have been reported for narrowband-selected LAEs at rest-frame $\sim$8000~\AA\ and $z = 2.4$--$4.5$ \citep{Im_et_al_2026}, from fits that exclude unresolved sources and allow a wider range of S\'ersic index. The exact values should be interpreted cautiously given the compact sizes and PSF sensitivity of the fits.

\begin{deluxetable*}{lccccc}
\tablecaption{LAE Size Properties by Narrowband\label{tab:size_mass}}
\tablehead{
\colhead{Narrowband} & \colhead{$R_{e,\mathrm{opt}}$ (kpc)} & \colhead{$R_{e,\mathrm{UV}}$ (kpc)} & \colhead{$R_{e,\mathrm{opt}}/R_{e,\mathrm{UV}}$} & \colhead{$\gamma$} & \colhead{$B$}
}
\startdata
NB387 & $0.91^{+0.07}_{-0.07}$ & $0.83^{+0.05}_{-0.05}$ & $1.05$ & $0.30 \pm 0.11$ & $-2.59 \pm 0.96$ \\
NB527 & $0.69^{+0.02}_{-0.02}$ & $0.57^{+0.04}_{-0.05}$ & $1.20$ & $0.14 \pm 0.18$ & $-1.36 \pm 1.56$ \\
NB718 & $0.50\pm0.03$ & $0.45^{+0.04}_{-0.04}$ & $1.12$ & $0.31 \pm 0.10$ & $-2.97 \pm 0.89$ \\
NB816 & $0.44^{+0.10}_{-0.04}$ & $0.31\pm0.04$ & $1.24$ & $0.17 \pm 0.14$ & $-1.79 \pm 1.19$ \\
NB921 & $0.52^{+0.13}_{-0.12}$ & $0.46^{+0.08}_{-0.02}$ & $1.05$ & $0.51 \pm 0.30$ & $-4.85 \pm 2.75$ \\
\enddata
\tablecomments{Redshifts and sample sizes for each narrowband are listed in Table~\ref{tab:sample}; NB973 ($z = 6.99$, $N = 4$) is excluded. Columns 2--3: median $R_e$ (semimajor axis) with 68\% bootstrap confidence intervals ($10{,}000$ iterations); rest-frame optical uses the filter covering rest-frame $\sim$4100--6400~\AA\ (varies by redshift), rest-frame UV uses $\lambda_\mathrm{rest} \sim 2200$--$3600$~\AA\ (see Table~\ref{tab:sample} for filter assignments). Column 4: median of the per-source ratio $R_{e,\mathrm{opt}}/R_{e,\mathrm{UV}}$. Columns 5--6: best-fit size--mass slope $\gamma$ and intercept $B$ from Equation~\ref{eq:size_mass} (Section~\ref{sec:size_mass}), with 68\% bootstrap uncertainties ($10{,}000$ iterations).}
\end{deluxetable*}

\subsection{Size--Mass Relation} \label{sec:size_mass}

\begin{figure*}
\includegraphics[width=\textwidth]{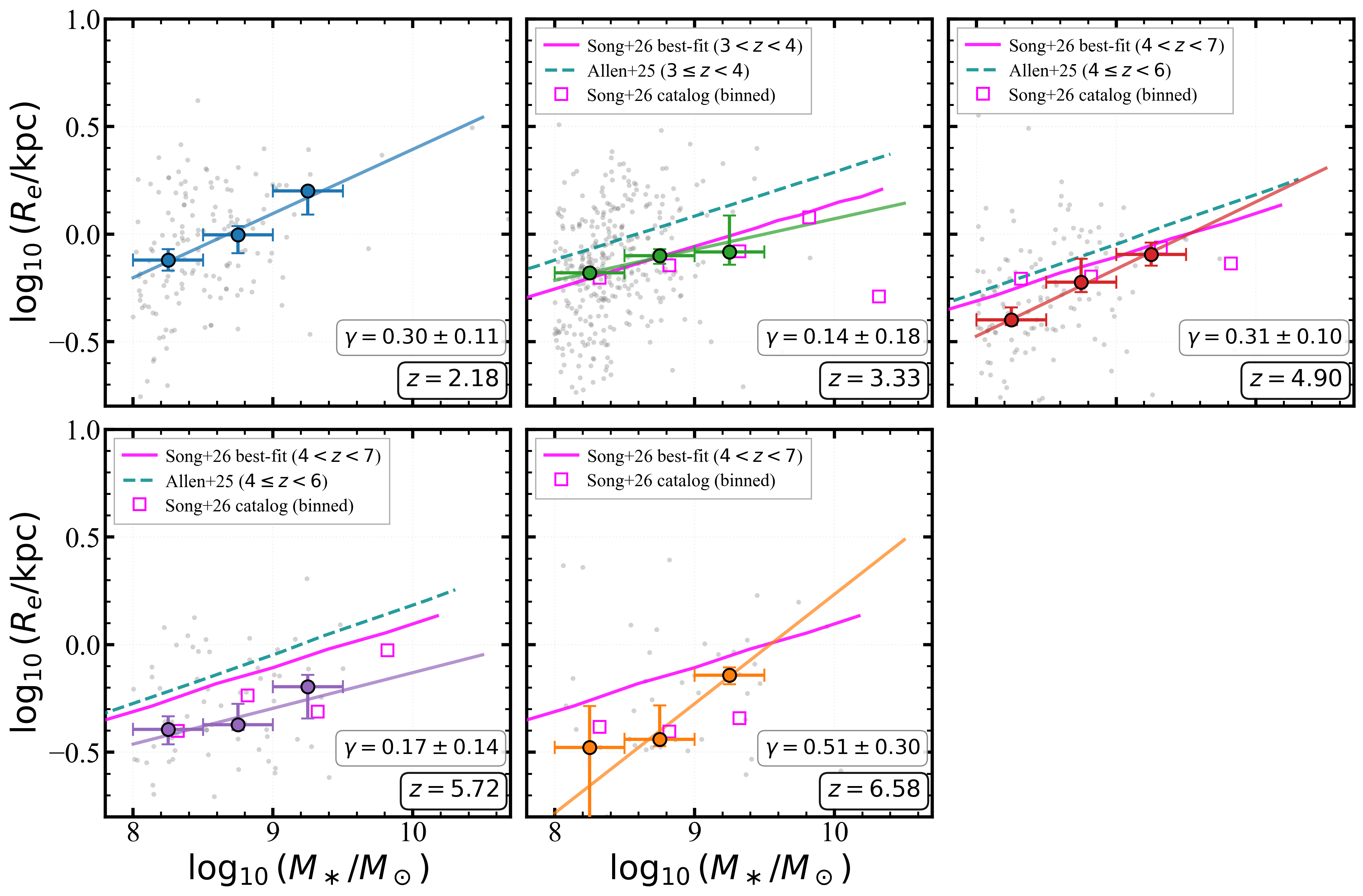}
\caption{Rest-frame optical size--mass relation for LAEs in five narrowband redshift bins ($z = 2.18$--$6.58$; NB973 at $z = 6.99$ is excluded owing to its small sample size). Filled circles with error bars show the binned median $\log_{10}(R_e/\mathrm{kpc})$ in 0.5~dex stellar-mass bins, with 68\% bootstrap confidence intervals ($10{,}000$ iterations). Gray points show individual sources. Solid colored lines show our best-fit size--mass relations (Equation~\ref{eq:size_mass}; slope $\gamma$ annotated in each panel). Where the redshift ranges overlap, solid magenta lines show the published LAE size--mass relations from \citet{Song_et_al_2026}, and open magenta squares show medians recomputed from their catalog in redshift windows matched to our narrowband bins. Dashed cyan lines show the SFG size--mass relations from \citet{Allen_et_al_2025}.}
\label{fig:size_mass}
\end{figure*}

We examine the relationship between rest-frame optical size and stellar mass for our LAE sample at each narrowband redshift. Following \citet{Song_et_al_2026}, we model the size--mass relation as
\begin{equation} \label{eq:size_mass}
\log_{10}\!\left(\frac{R_e}{\mathrm{kpc}}\right) = \gamma \, \log_{10}\!\left(\frac{M_\ast}{M_\odot}\right) + B,
\end{equation}
where $\gamma$ is the slope and $B$ is the intercept. For each narrowband, we compute binned medians in stellar mass bins of 0.5~dex width over $8.0 \leq \log(M_\ast/M_\odot) < 10.5$, and fit Equation~\ref{eq:size_mass} to the binned medians with uncertainties from $10{,}000$ bootstrap resamplings of individual sources. The results are presented in Figure~\ref{fig:size_mass} and Table~\ref{tab:size_mass}.

The measured slopes range from $\gamma = 0.14$ to $0.51$ across the five narrowband bins (Table~\ref{tab:size_mass}). While individual fits carry substantial uncertainties owing to the limited mass range (three valid bins spanning $8.0 \leq \log M_\ast < 9.5$ in most cases), the slopes are broadly consistent with $\gamma = 0.20 \pm 0.01$ at $3 < z < 4$ and $\gamma = 0.20 \pm 0.03$ at $4 < z < 7$ reported by \citet{Song_et_al_2026} for LAEs, as well as the rest-frame optical size--mass slopes of star-forming galaxies, $\gamma = 0.215 \pm 0.009$ averaged over $3 \leq z < 9$ \citep{Allen_et_al_2025}, $0.25^{+0.05}_{-0.03}$ at $3 < z < 5.5$ \citep{Ward_et_al_2024}, and $0.10$--$0.20$ at $2 < z < 8$ \citep{Yang_et_al_2025}.

Although the slopes broadly agree within uncertainties, the slope at $z=6.58$ ($\gamma=0.51\pm0.30$) is only weakly constrained given the small sample size ($N=51$). The normalization of the size--mass relation decreases toward higher redshift (Figure~\ref{fig:size_mass}). At $z = 3.33$ (NB527), our binned medians agree well with the \citet{Song_et_al_2026} $3 < z < 4$ fit. At $z \geq 4$, our data lie below the \citet{Song_et_al_2026} broad-bin $4 < z < 7$ fit (solid magenta lines in Figure~\ref{fig:size_mass}), with the offset appearing larger toward higher redshift. This apparent offset is reduced when the comparison is made in narrower redshift intervals; we recompute binned medians from the \citet{Song_et_al_2026} catalog at redshifts matched to each narrowband (i.e., within the redshift selection window of each narrowband filter) (open squares in Figure~\ref{fig:size_mass}) to illustrate this. At $z = 5.72$ (NB816), the redshift-matched \citet{Song_et_al_2026} medians are $\sim$0.1--0.2~dex lower than their broad-bin fit, bringing them into closer agreement with our data. At $z = 6.58$ (NB921), both our data and the \citet{Song_et_al_2026} sources show comparably low normalizations, although the small sample sizes at this redshift result in substantial scatter. The \citet{Allen_et_al_2025} SFG relation ($4 \leq z < 6$), derived from $\sim$3500 photometrically selected star-forming galaxies with JWST/NIRCam sizes in CEERS and PRIMER, lies above both our LAE data and the redshift-matched \citet{Song_et_al_2026} LAE medians at $z \gtrsim 5$, consistent with LAEs being compact relative to the general SFG population at fixed stellar mass. Comparing our binned medians directly with the \citet{Allen_et_al_2025} fitting relations in matched redshift and filter bins, the LAE normalization falls $\sim$0.1--0.2~dex below the SFG relation at fixed stellar mass across $z \sim 3$--$6$. Differences in selection functions and stellar-mass estimates likely dominate the systematic uncertainties in this comparison, with smaller contributions from fitting methodology and redshift binning; the SFG comparison sample is drawn from broadband photometric catalogs with different completeness criteria, whereas our LAE sample is selected by Ly$\alpha$ narrowband excess and is therefore governed primarily by Ly$\alpha$ line flux, equivalent width, and narrowband transmission limits (Section~\ref{sec:sample_construction}). \citet{Im_et_al_2026} similarly find that narrowband-selected LAEs at $z = 2.4$--$4.5$ lie below the rest-frame optical size--mass relation of typical SFGs, with a smaller offset in the Horizon Run 5 simulation. At $z = 4.90$, protocluster LAEs in the Loktak overdensity \citep{Laishram_et_al_2026b} are offset by $+0.12$~dex in rest-optical size at fixed stellar mass relative to field LAEs (Section~\ref{sec:discussion_environment}).

\subsection{Size--Equivalent Width and Size--Luminosity Relations} \label{sec:size_ew}

\begin{figure*}
\includegraphics[width=0.49\textwidth]{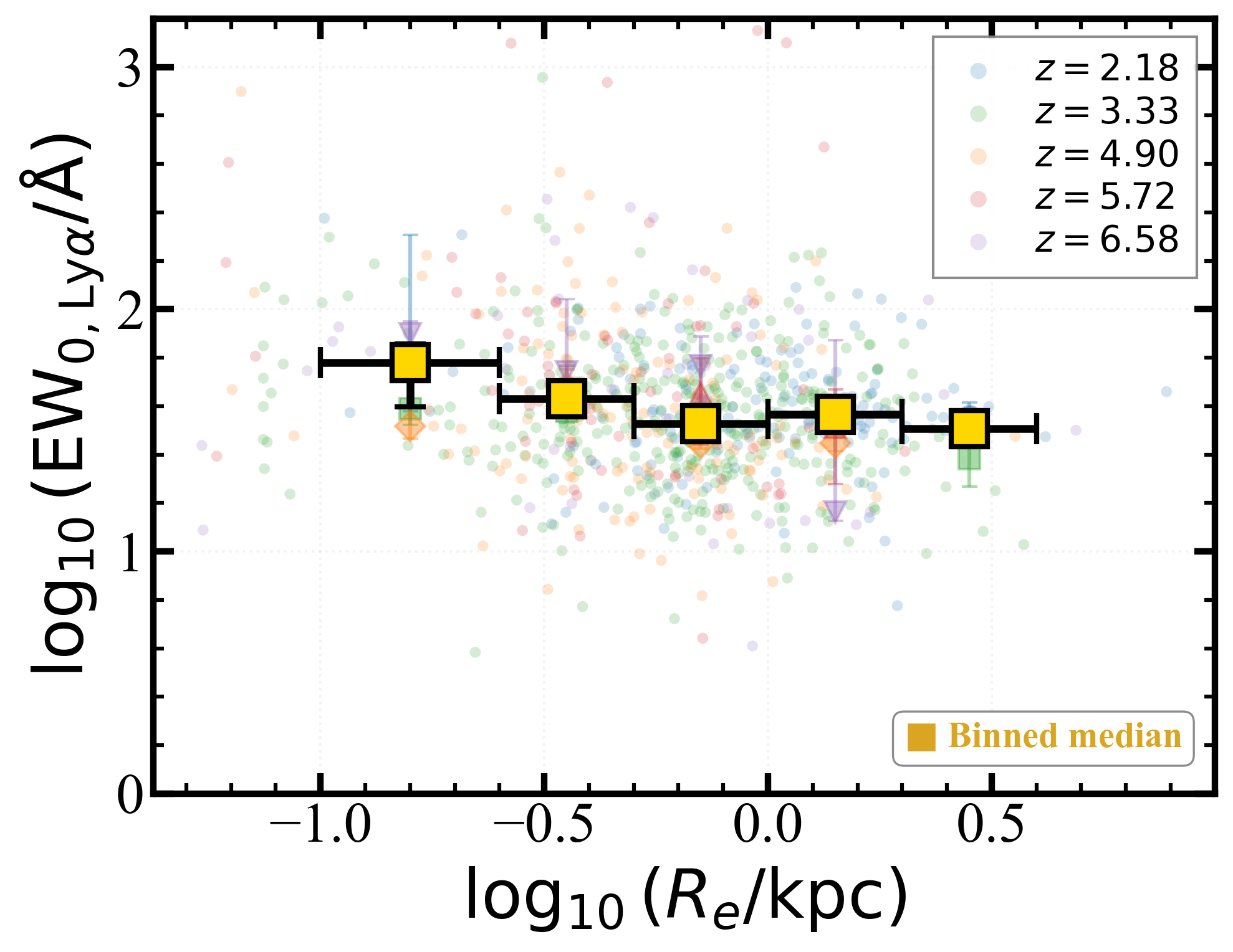}
\includegraphics[width=0.49\textwidth]{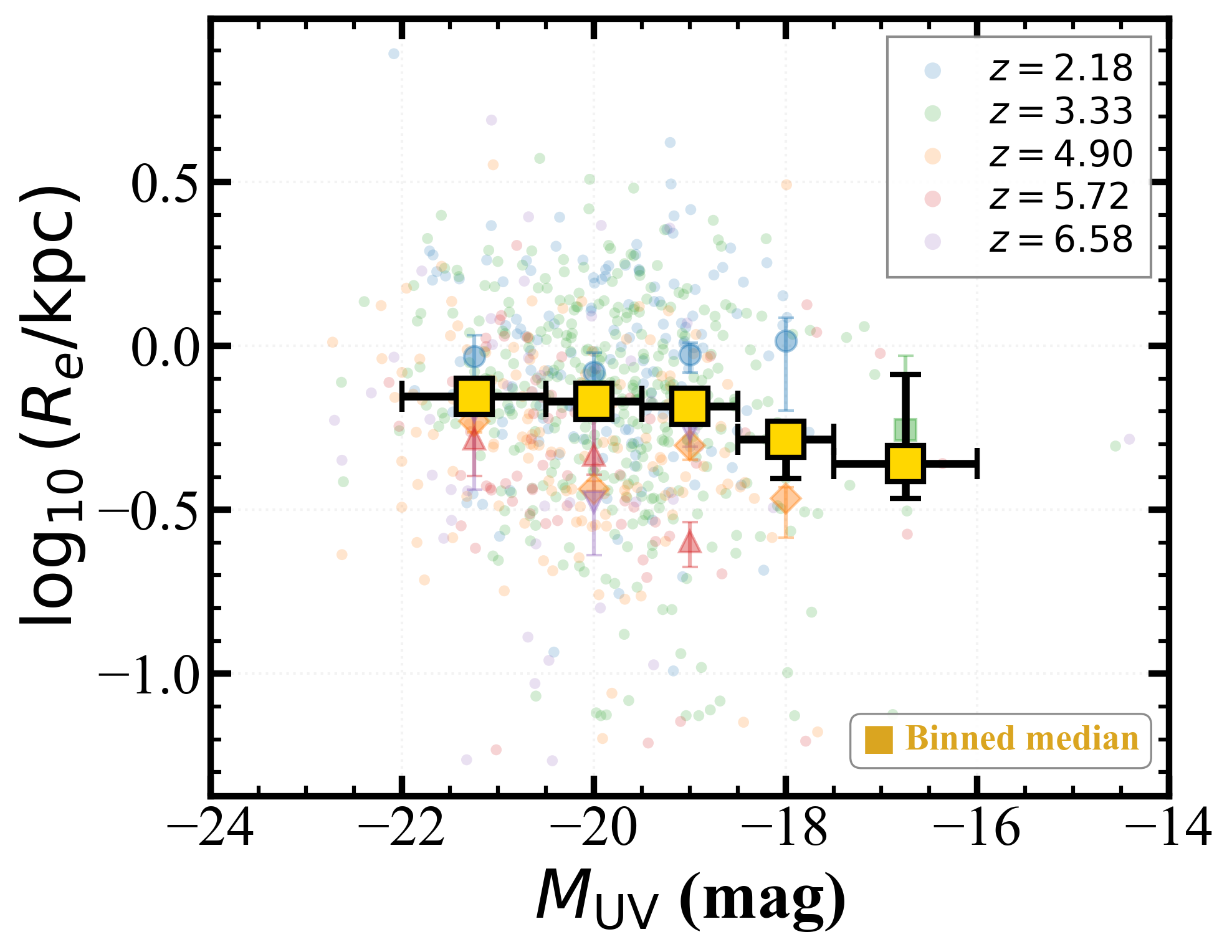}
\caption{\textit{Left:} Ly$\alpha$ rest-frame equivalent width as a function of rest-frame optical size for 766 LAEs across $z \sim 2$--$7$. \textit{Right:} Rest-frame optical size as a function of absolute UV magnitude for 753 LAEs. In both panels, colored points show individual sources at each narrowband redshift (see legend), faded symbols with error bars show per-narrowband binned medians, and gold squares show the combined binned medians with 68\% bootstrap confidence intervals ($10{,}000$ iterations). The left panel shows a weak but statistically significant anticorrelation between size and EW$_{0,\mathrm{Ly}\alpha}$ ($\rho_s = -0.17$, $p = 3.3 \times 10^{-6}$), while the right panel shows a weak but statistically significant combined size--luminosity correlation ($\rho_s = -0.09$, $p = 0.016$).}
\label{fig:size_ew_muv}
\end{figure*}

We investigate how the rest-frame optical size of LAEs relates to their Ly$\alpha$ equivalent width and UV luminosity. Figure~\ref{fig:size_ew_muv} presents $\mathrm{EW}_{0,\mathrm{Ly}\alpha}$ as a function of $R_e$ (left panel) and $R_e$ as a function of $M_\mathrm{UV}$ (right panel) for the combined sample across all narrowband redshifts.

A Spearman rank test on the combined sample of 766 LAEs yields $\rho_s = -0.167$ ($p = 3.3 \times 10^{-6}$), indicating a weak but statistically significant tendency for more compact LAEs to have higher $\mathrm{EW}_{0,\mathrm{Ly}\alpha}$. The binned medians in the left panel of Figure~\ref{fig:size_ew_muv} show that the smallest LAEs ($\log R_e \sim -0.8$) have a median $\log(\mathrm{EW}_0\,[\text{\AA}]) \approx 1.78$ ($\mathrm{EW}_0 \approx 60$~\AA), while the largest ($\log R_e \sim 0.45$) have $\log(\mathrm{EW}_0\,[\text{\AA}]) \approx 1.51$ ($\mathrm{EW}_0 \approx 32$~\AA). The strength of this anticorrelation varies with redshift: at $z < 4$, the correlation is weak ($\rho_s = -0.090$, $p = 0.040$), whereas at $z \geq 4$ the anticorrelation is stronger ($\rho_s = -0.241$, $p = 1.1 \times 10^{-4}$). The per-narrowband Spearman coefficients are $\rho_s = -0.062$ ($p = 0.44$) at $z = 2.18$, $-0.130$ ($p = 0.013$) at $z = 3.33$, $-0.227$ ($p = 0.009$) at $z = 4.90$, $-0.243$ ($p = 0.044$) at $z = 5.72$, and $-0.252$ ($p = 0.075$) at $z = 6.58$. This high-redshift signal is most pronounced in NB718 ($z = 4.90$), which is the most statistically significant individual bin; NB816 ($z = 5.72$) is marginally significant ($\rho_s = -0.243$, $p = 0.044$), while NB921 ($z = 6.58$) does not reach significance individually owing to its small sample size ($N = 51$). The apparent redshift dependence should therefore be interpreted cautiously. The anticorrelation is driven mainly by the most compact LAEs: 25\% of the sample have $R_e$ below half the PSF FWHM of the fitted filter (i.e., are marginally resolved), and these sources tend to have higher EW$_0$. The combined anticorrelation remains significant after excluding sources with $R_e$ below a quarter of the PSF FWHM ($\rho_s = -0.12$, $p = 0.001$), but not after excluding those below half the PSF FWHM ($\rho_s = -0.06$, $p = 0.17$). Excluding the 26 LAEs whose EW$_0$ is a lower bound (Section~\ref{sec:ew}) gives $\rho_s = -0.14$ ($p = 1.1 \times 10^{-4}$).

This size--EW anticorrelation is broadly consistent with the rest-frame UV result of \citet{PaulinoAfonso_et_al_2018}, who reported smaller UV sizes at higher EW$_0$ for LAEs at $z \sim 2$--$6$, and with \citet{Kerutt_et_al_2022}, who confirmed a similar anticorrelation ($p = 6.4 \times 10^{-16}$, Spearman rank test) for ${\sim}2000$ MUSE LAEs in rest-frame UV. Gravitationally lensed LAEs at $z = 1.7$--$3.3$ show a tentative anticorrelation between rest-frame UV size and Ly$\alpha$ EW \citep{Kim_et_al_2026}, although the sample selection, rest-frame wavelength, and size definitions differ. Similar trends are found for the rest-frame optical sizes of spectroscopically confirmed \citep{Song_et_al_2026} and narrowband-selected \citep{Im_et_al_2026} LAEs with JWST, whereas \citet{Bond_et_al_2012} found no evidence for a trend between EW and half-light radius in their HST rest-frame UV LAE samples.

The right panel of Figure~\ref{fig:size_ew_muv} shows that the combined size--luminosity relation for 753 LAEs with reliable $M_{1500}$ (13 sources lack a reliable UV magnitude) is statistically significant ($\rho_s = -0.088$, $p = 0.016$) but weaker than the size--EW anticorrelation. The per-narrowband Spearman coefficients for the size--luminosity relation are $\rho_s = -0.023$ ($p = 0.77$) at $z = 2.18$, $-0.189$ ($p = 3.2 \times 10^{-4}$) at $z = 3.33$, $-0.222$ ($p = 0.011$) at $z = 4.90$, $-0.213$ ($p = 0.086$) at $z = 5.72$, and $-0.077$ ($p = 0.62$) at $z = 6.58$. Within individual narrowband bins, NB527 ($z = 3.33$; $\rho_s = -0.189$, $p = 3.2 \times 10^{-4}$) and NB718 ($z = 4.90$; $\rho_s = -0.222$, $p = 0.011$) show significant size--luminosity correlations, with more UV-luminous LAEs tending to have larger $R_e$; no significant trend is detected in the remaining bins. These per-redshift trends are consistent with the size--luminosity relation reported by \citet{Shibuya_et_al_2019}, who found that LAEs follow an $R_e$--$L_\mathrm{UV}$ scaling broadly similar to those of SFGs and LBGs. Combining LAEs across different redshifts, each with a different size normalization from cosmic size evolution (Section~\ref{sec:size_evolution}), may dilute the per-epoch trends and produce a weaker combined signal. Because $\mathrm{EW}_{0,\mathrm{Ly}\alpha}$ and UV luminosity are not independent quantities (fainter UV continua at fixed Ly$\alpha$ flux yield higher EW), we compute the partial Spearman correlation between $R_e$ and $\mathrm{EW}_{0,\mathrm{Ly}\alpha}$ controlling for $M_\mathrm{UV}$. The partial correlation remains significant ($\rho_\mathrm{partial} = -0.126$, $p = 5.5 \times 10^{-4}$, $N = 753$), suggesting that the size--EW anticorrelation is not solely driven by the size--luminosity relation. In the sample of \citet{Im_et_al_2026}, the Ly$\alpha$ EW also correlates strongly with stellar mass, whereas in our sample the size--EW anticorrelation remains significant after controlling for stellar mass ($\rho_\mathrm{partial} = -0.155$, $p = 1.6 \times 10^{-5}$, $N = 766$).

\subsection{Dependence of LAE Sizes on Environment} \label{sec:environment}

We examine whether rest-frame optical LAE sizes depend on local environment using the fifth-nearest-neighbor overdensity $\delta_\mathrm{NN}$ described in Section~\ref{sec:env_method}.

\begin{figure*}
\includegraphics[width=\textwidth]{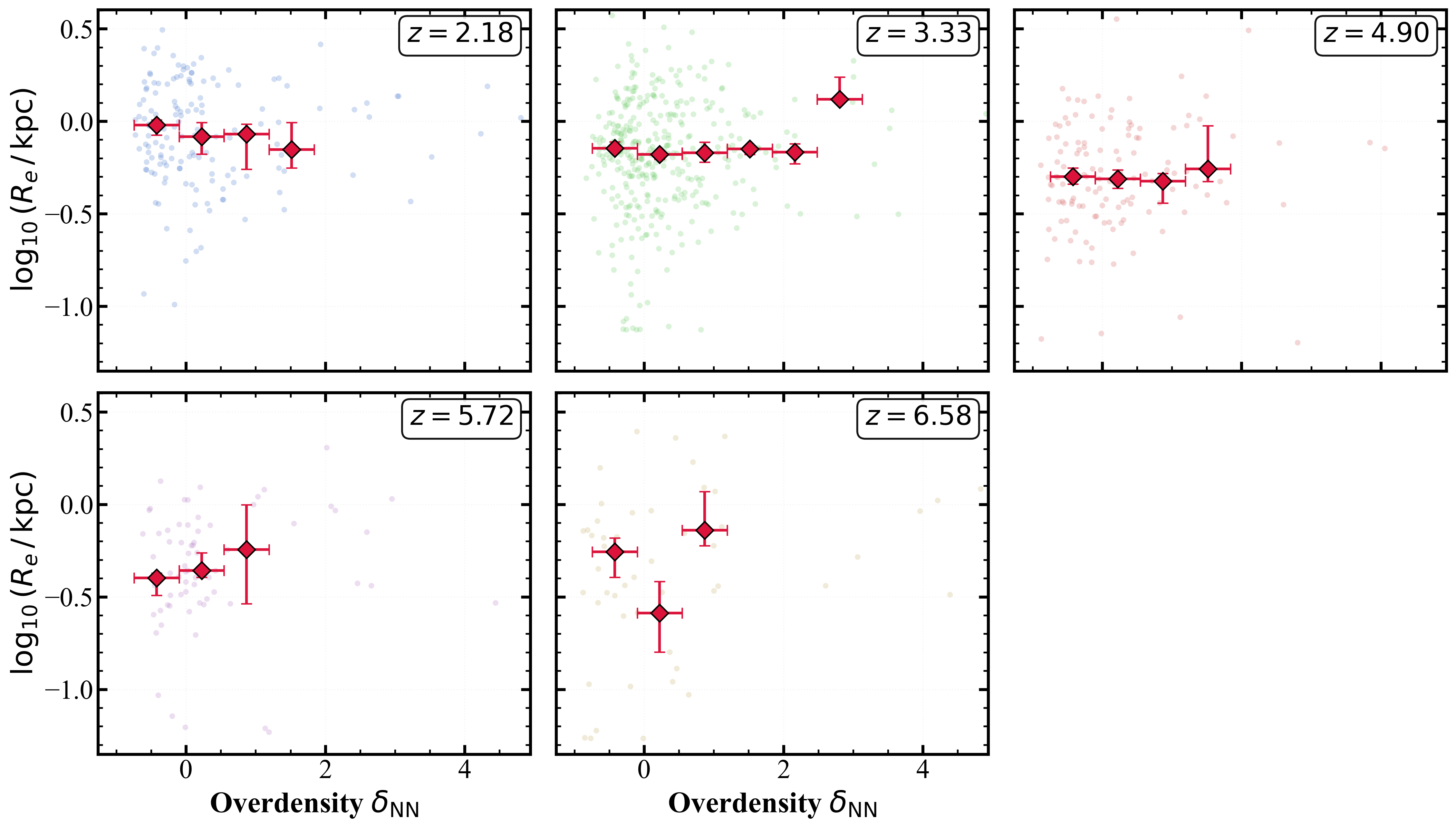}
\caption{Rest-frame optical effective radius $R_e$ as a function of the fifth-nearest-neighbor overdensity $\delta_\mathrm{NN}$ for LAEs across five narrowband redshift bins (labeled in each panel). Small colored circles show individual LAEs; red filled diamonds with error bars indicate the median $\log_{10}(R_e/\mathrm{kpc})$ and 68\% bootstrap confidence intervals ($10{,}000$ iterations) computed in bins of $\delta_\mathrm{NN}$ of equal width. No statistically significant correlation between $R_e$ and $\delta_\mathrm{NN}$ is detected in any individual redshift bin ($p > 0.05$ in all cases; see Section~\ref{sec:environment}), and the combined-sample Spearman correlation is $\rho_s = +0.05$ ($p = 0.18$). The binned medians are consistent with a flat trend across the sampled range of $\delta_\mathrm{NN}$ at all redshifts.}
\label{fig:env_size}
\end{figure*}

Figure~\ref{fig:env_size} shows $\log_{10}(R_e/\mathrm{kpc})$ as a function of $\delta_\mathrm{NN}$ for each of the five narrowband subsamples. No statistically significant correlation is detected in any individual redshift bin: the Spearman rank correlation coefficients are $\rho_s = -0.07$ ($p = 0.36$) at $z = 2.18$, $+0.04$ ($p = 0.49$) at $z = 3.33$, $+0.11$ ($p = 0.20$) at $z = 4.90$, $+0.13$ ($p = 0.29$) at $z = 5.72$, and $+0.24$ ($p = 0.095$) at $z = 6.58$. Combining all five narrowband subsamples yields $\rho_s = +0.05$ ($p = 0.18$). The binned medians overplotted in each panel are consistent with a flat trend across the sampled range of $\delta_\mathrm{NN}$ in all redshift bins. The largest individual Spearman coefficient is found at $z = 6.58$ ($\rho_s = +0.24$, $p = 0.095$), which does not reach statistical significance given the limited sample size at this redshift.

We assess this result using three additional checks. First, we repeat the analysis with the Voronoi tessellation overdensity, $\delta_\mathrm{Vor}$, as an alternative environment estimator. No significant correlation is found in any narrowband ($p > 0.05$ in all bins), and the combined-sample Spearman correlation is $\rho_s = +0.03$ ($p = 0.48$), consistent with the $\delta_\mathrm{NN}$ result. Second, we test whether the absence of a size--environment signal could be affected by systematic variation of stellar mass or S\'ersic index with environment. Neither stellar mass ($\rho_s = -0.04$, $p = 0.29$) nor S\'ersic index ($\rho_s = 0.00$, $p = 0.92$) shows a significant correlation with $\delta_\mathrm{NN}$ in the combined sample. Third, we compare the highest- and lowest-density quartiles of $\delta_\mathrm{NN}$ within each narrowband subsample, which is more sensitive to a signal confined to the densest regions than a monotonic rank correlation. The highest-density quartile is nominally larger at $z = 4.90$, the redshift of the Loktak protocluster \citep{Laishram_et_al_2026b} ($+0.070$~dex; Mann-Whitney $p = 0.293$), and comparable at $z = 3.33$ ($+0.007$~dex; $p = 0.584$). No subsample shows a statistically significant difference ($p = 0.08$--$0.58$); the larger nominal differences at $z = 5.72$ and $6.58$ ($+0.19$ and $+0.23$~dex) are based on small subsamples.

We therefore find no significant correlation of LAE rest-frame optical size with the projected LAE-traced environment over the redshift range $z \simeq 2$--$7$ probed by our sample. This null result is unchanged when using either the fifth-nearest-neighbor or Voronoi-based overdensity estimator and persists in a comparison of the highest- and lowest-density quartiles; neither stellar mass nor S\'ersic index correlates significantly with $\delta_\mathrm{NN}$ in the combined sample. Because the overdensity is measured in projected narrowband slices, any three-dimensional environmental dependence on smaller scales may be diluted by projection effects; possible biases in LAE-traced density are discussed in Section~\ref{sec:discussion_environment}.

\section{Discussion} \label{sec:discussion}

\subsection{LAE Structural Evolution in Context} \label{sec:discussion_context}

Our rest-frame optical size evolution over $z \sim 2$--$7$ ($\beta_\mathrm{opt} = -0.94 \pm 0.09$) is in close agreement with the $V$-band result of \citet{Song_et_al_2026} at $3 \lesssim z < 7$ ($\beta_V = -0.93 \pm 0.18$), derived from an independent spectroscopically confirmed sample spanning multiple extragalactic fields. The agreement is notable given the different selection methods (narrowband versus spectroscopic) and field coverage (COSMOS versus multi-field), and implies mild LAE size growth from $R_e \approx 0.5$~kpc at $z \sim 5$--$7$ to $\sim$0.9~kpc at $z \sim 2$.

At $z \sim 2$--$3$, LAEs are a factor of $\sim$2--4 smaller than typical SFGs at similar UV luminosities \citep{PaulinoAfonso_et_al_2018, Kim_et_al_2026}, with this size difference decreasing toward $z \gtrsim 5$ as the general SFG population becomes more compact \citep{PaulinoAfonso_et_al_2018, Song_et_al_2026}. In the rest-frame optical size--mass comparison, our LAEs lie below the \citet{Allen_et_al_2025} SFG size--mass relation over the $z \sim 3$--$6$ redshift range considered here, with this offset persisting to $z \gtrsim 5$ (Section~\ref{sec:size_mass}). \citet{Song_et_al_2026} find that their spectroscopically confirmed LAEs closely match the \citet{Allen_et_al_2025} SFG normalization at $4 < z < 7$, though at $3 < z < 4$ their LAEs lie $\sim$0.1~dex below; our narrowband LAEs show a somewhat larger offset ($\sim$0.1--0.2~dex) across $z \sim 3$--$6$, suggesting that the degree of offset may vary with both redshift and sample selection. The decreasing UV size difference between LAEs and SFGs toward earlier epochs is consistent with the picture in which, as typical galaxy sizes were smaller at $z \gtrsim 5$, a larger fraction of the SFG population approached the compactness that may facilitate Ly$\alpha$ escape \citep{PaulinoAfonso_et_al_2018}, contributing to the observed rise in the Ly$\alpha$-emitting fraction toward $z \sim 6$. \citet{Shimizu_et_al_2025} found that the Ly$\alpha$ escape fraction remains approximately constant ($f_\mathrm{esc,Ly\alpha} \sim 40\%$) for young ($<$100~Myr) LAEs over $z = 2.2$--$6.6$, and argued that LAEs alone could supply sufficient ionizing photons for reionization at $z \sim 6$.

The modest UV--optical size offset ($\log(R_{e,\mathrm{opt}}/R_{e,\mathrm{UV}}) = 0.056^{+0.005}_{-0.006}$), somewhat larger than, but compatible with, the $\sim$0.03~dex from \citet{Song_et_al_2026}, indicates that the rest-frame UV and optical light distributions of LAEs are qualitatively similar across $z \sim 2$--$7$. We interpret this offset primarily as a morphological $K$-correction (Section~\ref{sec:wavelength_dependence}), consistent with the conclusion of \citet{Song_et_al_2026} that LAEs exhibit weak or negligible UV--optical color gradients. Combined with low S\'ersic indices throughout (median $n < 1.2$ at all redshifts; Section~\ref{sec:size_evolution}), comparable to or lower than the weighted mean $\langle n \rangle \approx 1.16$ reported for LAEs by \citet{Shibuya_et_al_2019}, these properties suggest that LAEs at $z \sim 2$--$7$ are compact, structurally simple systems with low-concentration, sub-exponential light profiles rather than bulge-dominated morphologies. A low S\'ersic index alone does not imply a rotationally supported disk: in the THESAN simulations at $z \simeq 6$--$8$, galaxies with $10^8 \leq M_\ast/M_\odot < 10^9$, the stellar-mass range containing most of our LAEs, are better described by a spherical shell model regulated by feedback-driven outflows and cold inflows than by disk formation, although this result is based on intrinsic stellar half-mass radii rather than light profiles \citep{Shen_et_al_2024}. Stacked rest-UV surface-brightness profiles of LAEs at $z = 2.84$ are likewise well described by a single exponential profile in most subsamples \citep{Kikuta_et_al_2023b}, with an additional extended component required only for the UV- and Ly$\alpha$-brightest and lowest-EW subsamples. The physical connection between compact morphology and Ly$\alpha$ emission is discussed in Section~\ref{sec:discussion_sfr_sd}.

\subsection{Star Formation Rate Surface Density and Ly$\alpha$ Escape} \label{sec:discussion_sfr_sd}

\begin{figure}
\includegraphics[width=0.99\columnwidth]{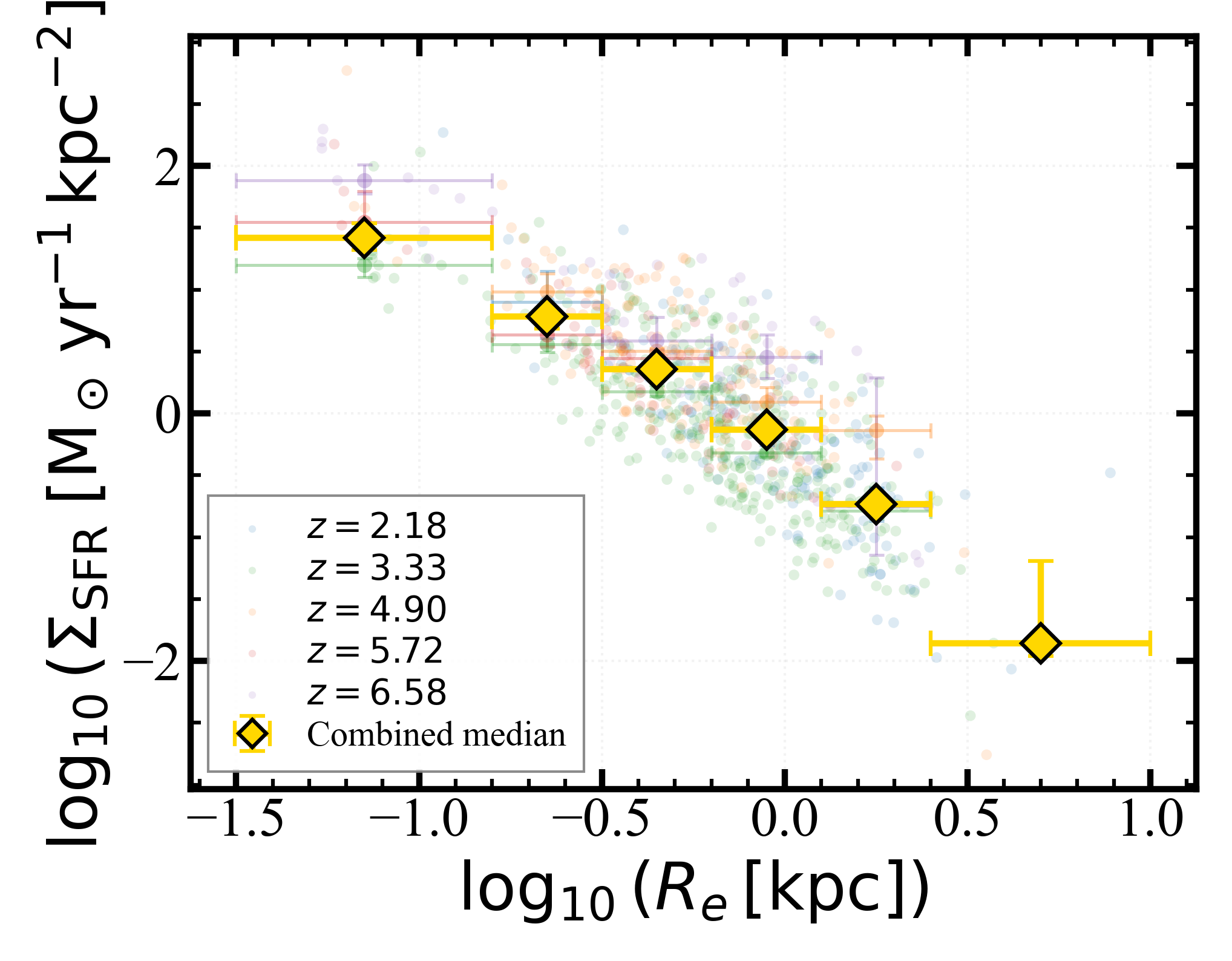}
\caption{Star formation rate surface density $\Sigma_\mathrm{SFR}$ (Equation~\ref{eq:sigma_sfr}) vs.\ rest-frame optical $R_e$ for 766 LAEs across $z \sim 2$--$7$, using dust-corrected SFR from LePhare SED fits. Colored points correspond to each narrowband redshift; gold squares show combined binned medians with 68\% bootstrap confidence intervals ($10{,}000$ iterations). The Spearman correlation is $\rho_s = -0.77$ ($p = 1.7 \times 10^{-151}$), but this anticorrelation is partly expected by construction because $\Sigma_\mathrm{SFR}$ depends explicitly on $R_e^{-2}$.}
\label{fig:sigma_sfr}
\end{figure}

The size--EW anticorrelation (Section~\ref{sec:size_ew}) may indicate that compact morphology is associated with conditions that facilitate Ly$\alpha$ emission, a relation also reported in rest-frame UV \citep{PaulinoAfonso_et_al_2018, Kerutt_et_al_2022, Kim_et_al_2026} and rest-frame optical \citep{Song_et_al_2026, Im_et_al_2026} studies. We quantify the concentration of star formation through the SFR surface density, following \citet{Shibuya_et_al_2019}:
\begin{equation} \label{eq:sigma_sfr}
\Sigma_\mathrm{SFR} = \frac{\mathrm{SFR}}{2\pi\,R_e^2} \quad [M_\odot\,\mathrm{yr}^{-1}\,\mathrm{kpc}^{-2}],
\end{equation}
where SFR is the dust-corrected value from LePhare SED fits \citep{Shuntov_et_al_2025} and $R_e$ is the rest-frame optical effective radius (semimajor axis). Because the SFRs are SED-based and $R_e$ is measured from broadband rest-frame optical light, $\Sigma_\mathrm{SFR}$ should be interpreted as an approximate global diagnostic rather than a spatially resolved SFR surface density, and is most uncertain for LAEs with sizes comparable to the PSF. Figure~\ref{fig:sigma_sfr} shows $\Sigma_\mathrm{SFR}$ versus $R_e$ for 766 LAEs. The anticorrelation is strong ($\rho_s = -0.77$, $p = 1.7 \times 10^{-151}$), spanning $>$3~dex from $\log\Sigma_\mathrm{SFR} \approx 1.4$ at $\log R_e < -0.8$ to $\approx -1.9$ at $\log R_e > 0.4$. However, this anticorrelation is partly expected by construction, since $\Sigma_\mathrm{SFR} \propto R_e^{-2}$, and the Spearman coefficient therefore does not by itself indicate a physical relationship. We therefore use $\Sigma_\mathrm{SFR}$ primarily as a descriptive quantity to assess whether compact LAEs occupy surface-density regimes associated with strong stellar feedback.

\citet{Kim_et_al_2026} argued that elevated $\Sigma_\mathrm{SFR}$ in compact LAEs may drive outflows that reduce the neutral-gas covering fraction or open low-column-density channels, facilitating Ly$\alpha$ escape, finding that 96\% of compiled LAEs (87 out of 91) exceed $\Sigma_\mathrm{SFR} \geq 1~M_\odot\,\mathrm{yr}^{-1}\,\mathrm{kpc}^{-2}$ using the same geometric form as Equation~\ref{eq:sigma_sfr}, but with circularized rest-frame UV radii rather than rest-frame optical semimajor-axis radii. With our adopted definition, LAEs at $z \gtrsim 5$ reach median $\Sigma_\mathrm{SFR} \gtrsim 2~M_\odot\,\mathrm{yr}^{-1}\,\mathrm{kpc}^{-2}$, exceeding their empirical threshold, while at $z \sim 2$--$3$ our medians (${\sim}0.7$--$0.8~M_\odot\,\mathrm{yr}^{-1}\,\mathrm{kpc}^{-2}$) approach but do not reach it. All medians exceed the $0.1~M_\odot\,\mathrm{yr}^{-1}\,\mathrm{kpc}^{-2}$ threshold for strong galactic winds \citep{Heckman_2001}. Since our $R_e$ is the rest-frame optical semimajor axis whereas \citet{Kim_et_al_2026} used circularized rest-frame UV radii, our $\Sigma_\mathrm{SFR}$ values are likely conservative relative to the adopted size definition.

In this scenario, concentrated star formation ($R_e \lesssim 1$~kpc) may generate feedback capable of opening low-column-density channels through the ISM, facilitating Ly$\alpha$ escape; this is in line with the $\Sigma_\mathrm{SFR}$--outflow velocity correlations observed in local starbursts \citep[e.g.,][]{Heckman_et_al_2015}. Radiation-hydrodynamic simulations of star-forming clouds also show Ly$\alpha$ photons escaping through turbulence-generated low-column-density channels evacuated by radiative feedback \citep{Kakiichi_Gronke_2021}, and in a cosmological zoom-in simulation at $z = 5$--$7$, sightlines with the highest Ly$\alpha$ equivalent widths are associated with outflowing regions of a starburst \citep{Smith_et_al_2019}. High $\Sigma_\mathrm{SFR}$ alone, however, may not suffice: starburst age, ISM geometry, and dust content modulate Ly$\alpha$ escape, and photoionization may also contribute \citep{Kim_et_al_2026}. \citet{Im_et_al_2026} find that the Ly$\alpha$ EW of narrowband-selected LAEs at $z = 2.4$--$4.5$ correlates strongly with the ratio of instantaneous to 100~Myr-averaged SFR, indicating a link to recent starburst activity. The apparent strengthening of the size--EW anticorrelation at $z \geq 4$ is most prominent in NB718 ($z = 4.90$) and marginally detected in NB816 ($z = 5.72$), and should be interpreted with caution; its physical origin remains unconstrained. One possible contributing factor is IGM attenuation at $z \gtrsim 4$, which reduces the observed Ly$\alpha$ flux and may alter the EW distribution of narrowband-selected LAEs; larger samples spanning multiple narrowband bins at $z > 4$ are needed to distinguish this selection effect from an intrinsic size--EW evolution.

\subsection{Dependence of LAE Morphology on Environment} \label{sec:discussion_environment}

The null correlation between LAE rest-frame optical size and local projected density (Section~\ref{sec:environment}) is consistent with the interpretation discussed in Section~\ref{sec:discussion_sfr_sd}. If Ly$\alpha$ escape is strongly influenced by $\Sigma_\mathrm{SFR}$-driven outflows that reduce the neutral-gas covering fraction \citep{Kim_et_al_2026, Shimizu_et_al_2025,Shimizu_et_al_2026}, then the relevant physical drivers may be largely internal to the galaxy and only weakly coupled to the surrounding projected density field. The additional finding that neither stellar mass nor S\'ersic index shows a significant correlation with $\delta_\mathrm{NN}$ (Section~\ref{sec:environment}) is compatible with this possibility.

This result is consistent with \citet{Malavasi_et_al_2021}, who found that the Ly$\alpha$ luminosity function and equivalent width distribution of LAEs in a $z \approx 3.78$ protocluster are comparable to field measurements at similar redshift. Together with our null size--density result, this suggests that LAE emission properties and rest-frame optical morphology show no strong correlation with projected large-scale environment in the regimes probed by these studies.

A targeted study of the Loktak LAE protocluster at $z = 4.90$ \citep{Laishram_et_al_2026b} reported a marginal rest-frame optical size enhancement among protocluster members (${\sim}40\%$ larger in rest-optical; Mann-Whitney $p = 0.041$). This is not in tension with the present null result: the Loktak analysis targeted an extreme overdensity (${\sim}4{\times}$ surface density enhancement within 1.5~pMpc), whereas the Spearman test here probes any monotonic trend across the full density range sampled by $\delta_\mathrm{NN}$, which includes many field and moderate-density sources. Rest-frame optical size enhancements in LAEs, if present, may therefore be confined to the most extreme protocluster cores. A similar pattern is seen in the circumgalactic gas: \citet{Kikuta_et_al_2023b} found that Ly$\alpha$ halo profiles of LAEs at $z = 2.84$ show little dependence on projected LAE overdensity outside the protocluster, whereas LAEs in the protocluster core show Ly$\alpha$ emission extending beyond 100~pkpc. Taken together with our result, this suggests that the weak environmental dependence extends from the stellar body of LAEs to their circumgalactic gas, with the possible exception of the most extreme overdensities. Both studies rely on projected LAE-traced density measures and are therefore subject to the same projection effects. LAE-traced density is also subject to selection effects that act most strongly in dense regions. \citet{Shimakawa_et_al_2017} found few LAEs within the H$\alpha$-emitter overdensities of a $z = 2.53$ protocluster, together with systematically lower Ly$\alpha$ escape fractions for H$\alpha$ emitters in higher-density regions, and cautioned that LAE-based surveys may largely miss galaxies in the densest protocluster cores. \citet{Momose_et_al_2021} similarly reported that LAEs at $z \sim 2$ do not trace neutral hydrogen isotropically, with those lying behind dense gas preferentially undetected in Ly$\alpha$. Such biases may reduce our sensitivity to environmental trends in precisely the regimes where they are expected to be strongest.

Two limitations are worth noting. The $\delta_\mathrm{NN}$ estimator traces projected density within narrowband slices ($\Delta z \sim 0.05$--$0.1$, corresponding to $\sim$20--60~comoving Mpc), reducing sensitivity to compact overdensities where environmental effects are most pronounced. The subsamples at $z \geq 5.72$ are also small ($N \leq 69$), limiting statistical power at the highest redshifts. Deeper studies of confirmed protoclusters at $z \gtrsim 5$ with JWST/NIRSpec will provide a more stringent test of whether environment shapes LAE morphology during the epoch of reionization.

\section{Summary and Conclusions} \label{sec:conclusions}

We have studied the rest-frame UV and optical morphologies of 770 LAEs at $z = 2.18$--$6.99$ (766 at $z \leq 6.58$ in the quantitative analyses) using JWST NIRCam imaging from the COSMOS-Web treasury survey \citep{Casey_et_al_2023}, combined with the SILVERRUSH multi-narrowband LAE catalog \citep{Kikuta_et_al_2023}. S\'ersic profile fitting was performed in four NIRCam filters (F115W, F150W, F277W, and F444W) for LAEs spanning $\sim$2.2~Gyr of cosmic time. Our main findings are as follows.

(1) LAE effective radii show a mild but statistically significant increase from rest-frame UV to optical, with a median ratio of $R_{e,\mathrm{opt}}/R_{e,\mathrm{UV}} = 1.14^{+0.01}_{-0.02}$ ($\log(R_{e,\mathrm{opt}}/R_{e,\mathrm{UV}}) = 0.056^{+0.005}_{-0.006}$; Wilcoxon $p = 6.3 \times 10^{-19}$, $N = 766$). The offset is detected at $z \leq 5.72$ but not at $z = 6.58$ ($p = 0.46$, $N = 51$), consistent with \citet{Song_et_al_2026} and indicating qualitatively similar UV and optical structures across $z \sim 2$--$7$.

(2) The rest-frame optical size evolves as $R_e \propto (1+z)^{-0.94 \pm 0.09}$, with median $R_e$ decreasing from $0.91^{+0.07}_{-0.07}$~kpc at $z = 2.18$ to $0.44^{+0.10}_{-0.04}$~kpc at $z = 5.72$; the rest-frame UV follows a consistent slope ($\beta_\mathrm{UV} = -0.92 \pm 0.21$), in good agreement with \citet{Song_et_al_2026} ($\beta_V = -0.93 \pm 0.18$), while extending the measured evolution down to $z = 2.18$. The median S\'ersic index remains $n \lesssim 1.2$ at all redshifts, indicating low-concentration or sub-exponential profiles with no strong evidence for bulge-dominated morphologies.

(3) The rest-frame optical size--mass relation slopes ($\gamma = 0.14$--$0.51$) are broadly consistent with the LAE values reported by \citet{Song_et_al_2026} ($\gamma \approx 0.20$) and the general SFG population from \citet{Allen_et_al_2025} ($\gamma = 0.215 \pm 0.009$). In normalization, LAEs lie $\sim$0.1--0.2~dex below the \citet{Allen_et_al_2025} SFG size--mass relation at fixed stellar mass over $z \sim 3$--$6$, with the offset persisting to $z \gtrsim 5$.

(4) More compact LAEs tend to have higher Ly$\alpha$ equivalent widths ($\rho_s = -0.167$, $p = 3.3 \times 10^{-6}$, $N = 766$), a trend driven mainly by the most compact, marginally resolved sources, with the anticorrelation stronger at $z \geq 4$ ($\rho_s = -0.241$, $p = 1.1 \times 10^{-4}$). A weaker size--UV luminosity trend ($\rho_s = -0.088$, $p = 0.016$) suggests that the size--EW anticorrelation is unlikely to be explained solely by luminosity. Compact LAEs also occupy elevated $\Sigma_\mathrm{SFR}$ regimes (noting that $\Sigma_\mathrm{SFR} \propto R_e^{-2}$ by construction), suggesting that compact morphology and concentrated star formation may facilitate Ly$\alpha$ escape.

(5) No significant correlation of LAE rest-frame optical size with projected LAE-traced density is detected in any narrowband subsample or in the combined sample, using either the fifth-nearest-neighbor estimator ($\rho_s = +0.05$, $p = 0.18$, $N = 766$) or the Voronoi-based overdensity ($\rho_s = +0.03$, $p = 0.48$). This null result suggests that LAE morphology may be governed primarily by internal galaxy properties, with any environmental dependence being weak, projection-diluted, or confined to the most extreme overdensities \citep{Laishram_et_al_2026b}.

LAEs over $z \sim 2$--$7$ are thus compact ($R_e \lesssim 1$~kpc), low-S\'ersic-index systems undergoing mild size growth with decreasing redshift, with their Ly$\alpha$ emission possibly linked to compact morphology and concentrated star formation. The primary limitations are the use of projected density estimators, which reduce sensitivity to compact overdensities, limited statistical power at $z \geq 5.72$ ($N \leq 69$), and the dependence of the size--EW result on marginally resolved sources. Future wide-field spectroscopy with Subaru/PFS \citep{Greene_et_al_2022} and targeted studies of confirmed $z \gtrsim 5$ protoclusters with JWST/NIRSpec will enable more precise environment mapping and provide a direct test of whether rest-frame optical size enhancements in LAEs are confined to the most extreme overdensities.

\section{Data Availability}
The JWST COSMOS-Web imaging and photometric catalog used in this work were retrieved from the COSMOS2025 data release, available at \url{https://cosmos2025.iap.fr}. The multi-narrowband LAE catalogs from the SILVERRUSH survey \citep{Kikuta_et_al_2023} used in this work will be made publicly available following the HSC-SSP Public Data Release 4 (PDR4).

\begin{acknowledgments}
We thank Sang Hyeok Im for helpful discussions, and we are grateful to the COSMOS-Web team. This work was supported by JSPS KAKENHI grant Nos. 23H01219 and 23K25915 and JSPS Core-to-Core Program (grant Nos. JPJSCCA20210003 and JPJSCCA20260002). H.K. acknowledges support from JSPS KAKENHI grant Nos. 23KJ2148 and 25K17444. S.K. acknowledges support from JSPS KAKENHI grant Nos. 24KJ0058 and 24K17101. T.K. and Y.K. acknowledge financial support from JSPS KAKENHI Grant Numbers 24H00002 (Specially Promoted Research by T.\ Kodama et al.) and 22K21349 (International Leading Research by S.\ Miyazaki et al.). Y.K. also acknowledges support from JSPS KAKENHI grant Nos. 26H02070 and 26K00751. SS is supported by the Japan Society for the Promotion of Science (JSPS) KAKENHI grant number JP26KJ0916.

This research is based in part on data collected at the Subaru Telescope, which is operated by the National Astronomical Observatory of Japan. We are honored and grateful for the opportunity of observing the Universe from Maunakea, which has the cultural, historical, and natural significance in Hawaii. 

The Hyper Suprime-Cam (HSC) collaboration includes the astronomical communities of Japan and Taiwan, and Princeton University. The HSC instrumentation and software were developed by the National Astronomical Observatory of Japan (NAOJ), the Kavli Institute for the Physics and Mathematics of the Universe (Kavli IPMU), the University of Tokyo, the High Energy Accelerator Research Organization (KEK), the Academia Sinica Institute for Astronomy and Astrophysics in Taiwan (ASIAA), and Princeton University. Funding was contributed by the FIRST program from the Japanese Cabinet Office, the Ministry of Education, Culture, Sports, Science and Technology (MEXT), the Japan Society for the Promotion of Science (JSPS), Japan Science and Technology Agency (JST), the Toray Science Foundation, NAOJ, Kavli IPMU, KEK, ASIAA, and Princeton University.
The HSC data were retrieved from the HSC data archive system, which is operated by Subaru Telescope and Astronomy Data Center (ADC) at NAOJ.
\end{acknowledgments}


\bibliography{sample631}{}
\bibliographystyle{aasjournal}

\end{document}